\documentclass[letter,journal]{IEEEtran}
\usepackage[table,xcdraw]{xcolor}
\usepackage{color,soul}
\usepackage{cite}
  \usepackage{amsmath}

\ifCLASSINFOpdf
   \usepackage[pdftex]{graphicx}
  \graphicspath{{./Figures/}}
  \DeclareGraphicsExtensions{.pdf,.jpeg,.png}
\else
  \usepackage[dvips]{graphicx}
  \graphicspath{{./Figures/}}
  \DeclareGraphicsExtensions{.eps}
\fi

\usepackage{tikz}
\usetikzlibrary{arrows,shapes,chains,calc,patterns,decorations.pathreplacing}
\usepackage{comment}
\usepackage{amsthm,amsfonts,dsfont,bm}

\usepackage{enumitem}
\usepackage[linesnumbered,lined,ruled,norelsize]{algorithm2e}
\usepackage{algorithmic}
\SetAlFnt{\small}

\usepackage{array}
\usepackage{multirow}
\newcommand{\tabincell}[2]{\begin{tabular}{@{}#1@{}}#2\end{tabular}}  

\ifCLASSOPTIONcompsoc
  \usepackage[caption=false,font=normalsize,labelfont=sf,textfont=sf]{subfig}
\else
  \usepackage[caption=false,font=footnotesize]{subfig}
\fi

\usepackage{dblfloatfix}

\ifCLASSOPTIONcaptionsoff
  \usepackage[nomarkers]{endfloat}
 \let\MYoriglatexcaption\caption
 \renewcommand{\caption}[2][\relax]{\MYoriglatexcaption[#2]{#2}}
\fi

\usepackage{url}
\usepackage[framemethod=tikz]{mdframed}
\usepackage{hhline}
\usepackage{fancyhdr}
\usepackage{booktabs}
\usepackage{longtable}
\usepackage{mfirstuc}
\usepackage{amssymb}
\usepackage{mathabx}
\usepackage{graphicx}
\newcommand{\RR}{\mathbb{R}}
\newcommand{\dd}{\mathrm{d}}
\newcommand{\A}{\mathcal{A}}

\newcommand{\NN}{\mathbb{N}}

\newcommand{\J}{\mathcal{J}}

\newcommand{\K}{\mathrm{K}}

\newcommand{\LX}{\mathcal{L}}

\newcommand{\cH}{\mathcal{H}}
\newcommand{\PX}{\mathcal{P}}
\newcommand{\SX}{\mathcal{S}}

\newcommand{\GX}{\mathcal{G}}

\newcommand{\NX}{\mathcal{N}}

\newcommand{\VV}{\mathcal{V}}

\newcommand{\F}{\mathcal{F}}  %%% add year 2022
\newcommand{\Lrm}{\mathrm{L}}   %%% add year 2022
\newcommand{\srm}{\mathrm{s}}   %%% add year 2022
\definecolor{lightblue}{rgb}{0.9, 1, 1}
\definecolor{lightyellow}{rgb}{1, 1, 0.9}
\definecolor{lightpink}{rgb}{1, 0.9, 1}
\definecolor{lightgreen}{rgb}{0.9, 1, 0.9}
\definecolor{lightgray}{rgb}{0.66, 0.66, 0.66}
\newcommand{\lightred}[1]{\textcolor[rgb]{0.8745, 0.498, 0.498}{#1}} %%% add year 2022
\makeatletter
\newcommand{\removelatexerror}{\let\@latex@error\@gobble}
\newcommand*\bigcdot{\mathpalette\bigcdot@{.5}}
\newcommand*\bigcdot@[2]{\mathbin{\vcenter{\hbox{\scalebox{#2}{$\m@th#1\bullet$}}}}}
\makeatother
\begin{document}

\title{A Map-matching Algorithm with Extraction of Multi-group Information for Low-frequency Data} %hierarchical

\author{Jie~Fang,~\IEEEmembership{Member,~IEEE},~Xiongwei~Wu,~Dianchao~Lin$^{*}$,~\IEEEmembership{Member,~IEEE}, ~~ ~ ~~ ~~~ Mengyun~Xu$^{*}$,~Huahua~Wu,~Xuesong~Wu,~Ting~Bi,~\IEEEmembership{Member,~IEEE}
	\thanks{Jie~Fang, Xiongwei Wu, Huahua Wu and Xuesong Wu are with the College of Civil Engineering, Fuzhou University, Fuzhou 350108, China (e-mail: fangjie@fzu.edu.cn; 200520113@fzu.edu.cn; 200527227@fzu.edu.cn; xuesong.wu.6@gmail.com
		%DianChao Lin is with the School of Economics and Management, Fuzhou University, Fuzhou 350108, China (email: lindianchao@126.com)
	}% <-this % stops a space
	\thanks{$^*$Corresponding author. Dianchao Lin is with the School of Economics and Management, Fuzhou University, Fuzhou 350108, China (email: \\ {lindianchao@fzu.edu.cn})}
	%\thanks{$^{\dagger}$ Corresponding author. Li Li is with the School of Civil Engineering, Fuzhou University, Fuzhou 350108, China (email: livelikeflowers@126.com)}
	\thanks{$^*$Corresponding author. Mengyun Xu is with the Intelligent Transport System Research Center, Wuhan University of Technology, Wuhan 430063, China (e-mail: xmymymy@foxmail.com).}
	\thanks{Ting Bi is with Computer Science, Maynooth University, Ireland (e-mail: ting.bi@mu.ie}
}

\maketitle

\begin{abstract}
	The growing use of probe vehicles generates a huge number of GNSS data. Limited by the satellite positioning technology, further improving the accuracy of map-matching is challenging work, especially for low-frequency trajectories. When matching a trajectory, the ego vehicle's spatial-temporal information of the present trip is the most useful with the least amount of data. In addition, there are a large amount of other data, e.g., other vehicles' state and past prediction results, but %the concentration of their value is very low for current matching accuracy. 
	it is hard to extract useful information for matching maps and inferring paths.
	Most map-matching studies only used the ego vehicle's data and ignored other vehicles' data. %a small part of data with high concentration of value and ignored the data with low concentration of value. 
	Based on it, this paper designs a new map-matching method to make full use of ``Big data''. We first sort all data into four groups according to their spatial and temporal distance from the present matching probe which allows us to sort for their usefulness. % We first sort all data into four groups (three useful groups and one useless group). 
	Then we design three different methods to extract valuable information (scores) from them: a score for speed and bearing, a score for historical usage, and a score for traffic state using the spectral graph Markov neutral network. Finally, we use a modified top-K shortest-path method to search the candidate paths within an ellipse region and then use the fused score to infer the path (projected location). We test the proposed method against baseline algorithms using a real-world dataset in China. The results show that all scoring methods can enhance map-matching accuracy. Furthermore, our method outperforms the others, especially when GNSS probing frequency is $\bm{\leq 0.01}$ Hz.%very low. 
\end{abstract}


\IEEEpeerreviewmaketitle

\section{Introduction} 
\label{S:Intro}

\IEEEPARstart{W}ITH the development of intelligent transportation systems, application of Global Navigation Satellite System (GNSS, including GPS, Beidou, etc) data consisting of vehicle ID and spatial-temporal data has become commonplace. GNSS-based traffic information is now the premise and foundation of intelligent traffic research, such as travel time estimation \cite{xu2019utilizing,xu2022intelligent} and traffic speed prediction \cite{yu2020forecasting,cui2020graph,zhao2019t}. 
Matching the GNSS trajectory onto the map is usually the first step of GNSS-based research. So far, map matching (MM) algorithms have made great accuracy for high-frequency GNSS probes %(HFP) 
\cite{atia2017low,sharath2019dynamic,wu2020map}. 
However, high-frequency probes (e.g., $\ge$1 Hz) come along with high cost, energy consumption, computing performance and storage space. %On the contrary, low-frequency GNSS data (e.g., $\le 1/30$ Hz or $\ge 30$ s sampling interval) could further reduce the data storage and transmission costs, but 
That is the reason why GNSS equipment with low-frequency probes %(LFP) %acquisition 
is still widely used in fleet management, travel time estimation, a release of bus stop information, etc.
Compared with an algorithm based on high-frequency probes, the MM with low-frequency probes (e.g., $\le 1/30$ Hz or $\ge 30$ s sampling interval) usually has lower accuracy. Its accuracy further drops significantly with a decrease of its frequency when the sampling interval is $>60$ s \cite{song2018hidden,wu2020map,liu2020deep}. How to further improve the MM accuracy with low-frequency probes is still challenging work.

In recent years, some governments and companies continue to collect the GNSS data from urban traffic, whose amount is ``big'' and increases fast day by day. Although 
the number of GNSS data (and corresponding calculated results) is large, past studies usually only used a few probes in a matching.  %a high concentration of value (COV) for improving the accuracy of a given MM trail. 
On the contrary, a large amount of data seem unrelated to the present MM work but maybe also involve valuable information.  For example, a vehicle's historical data can reflect its traveling habit, neighboring vehicles with the same orientation-destination (OD) usually have similar path choices, and recent traffic state in the urban network may influence ego vehicle's path choice in the near future. If we take appropriate methods, we can extract information from them. 

Vehicle's present GNSS data (this probe and last probe) should be the most valuable for current MM, and many classical studies only used this information to match the location and path. The spatial features, including geometry (e.g., location and network’s “shape”) and the topology (e.g., roads’ connectivity and bearing), are most commonly used \cite{quddus2007current}, and many research matched trajectories based on spatial data alone \cite{he2013line,rahmani2013path,quddus2015shortest,wang2019bmi,luo2020incremental}. Furthermore, some MM algorithms combine the spatial features with the temporal features, such as travel time \cite{rahmani2017travel,li2017citywide} and speed \cite{alrassy2021obd,kang2017online}, and sometimes their results could be better. In general, present two GNSS probes are the basic information for all MM algorithms. %\hl{There are only a few amounts, but their prediction results are usually not bad.} %Its amount is the minimum, but its concentration of value is the highest%because using few data can attribute a lot to the improve of MM accuracy

The historical data can indicate valuable information.
Zheng \textit{et al.} leveraged historical spatial features for MM \cite{zheng2012reducing}, and many research further use historical temporal features to improve MM accuracy \cite{ozdemir2018hybrid,liu2020deep,li2021trajectory,zhang2021map}. In addition,
based on the historical path inference results, some studies estimated travelers' path preference and used this information to fix present MM result \cite{froehlich2008route,song2018hidden,xiao2021path,yu2022map}. 
Furthermore, a vehicle's MM can also benefit from its neighbors if they have the same OD \cite{zheng2012reducing,bian2020trajectory}. 
%Although OH-data contain some valuable information, such as preferred speed and path, it may be useless if the OD of vehicle's present trip is new. In addition, different traveling periods (e.g., morning peak hour and night time) may indicate different traveling choices, and this information can also exist in other vehicles' historical behaviors in C-data.
The data from them are called \emph{collaborative information}, but only a few of research have taken it into consideration. Bian et al. proposed a collaboration based MM method, resampling GNSS points around into collaborative GNSS points to supplement the missing location information (while same OD is not necessary) \cite{bian2020trajectory}. Such a collaboration is different from our collaborative method which fully considers the similarity of the whole trajectories.

Theoretically, traffic state estimation results can also contribute to MM accuracy, but it's still lack of study. Similar work is to use all recent vehicles' data to match (microscopic) trajectories, such as a data augmentation method for MM \cite{zhao2019deepmm}. Furthermore, many research found that, compared with traditional methods including Kalman filter and support vector regression, the tool of neural network (NN) usually performed the best in %the related traffic research
traffic state estimation \cite{ma2015long,li2017diffusion}. Therefore, this paper takes the NN framework to extract traffic state information from all recent data and use this information to improve MM accuracy. 
%Although it is popular to use A-data to estimate travel time and traffic speed on road segment level, only few studies used A-data 
Note that there are several kinds of NN methods that are suitable for road-segment-level traffic prediction, including artificial NN \cite{xu2019utilizing,lu2020information}, convolutional NN (CNN) \cite{dabiri2018inferring}, Recurrent NN (RNN, including long short-term memory NN) \cite{ma2015long,zhao2019deepmm}, graph NN (GNN) \cite{wang2020traffic}, etc. Actually, different NN methods are not mutually exclusive. For example, there are algorithms with GNN+CNN (GCN) \cite{zhao2019t,guo2019attention,yu2020forecasting}, CNN+RNN+GNN \cite{li2017diffusion} and CNN+GNN+Markov model (spectral graph Morkov NN, SGMN) \cite{cui2020graph}. %This paper picks the SGMN as our extraction tool.

\begin{figure}[h!]
	\centering
	\includegraphics[width=0.75\linewidth]{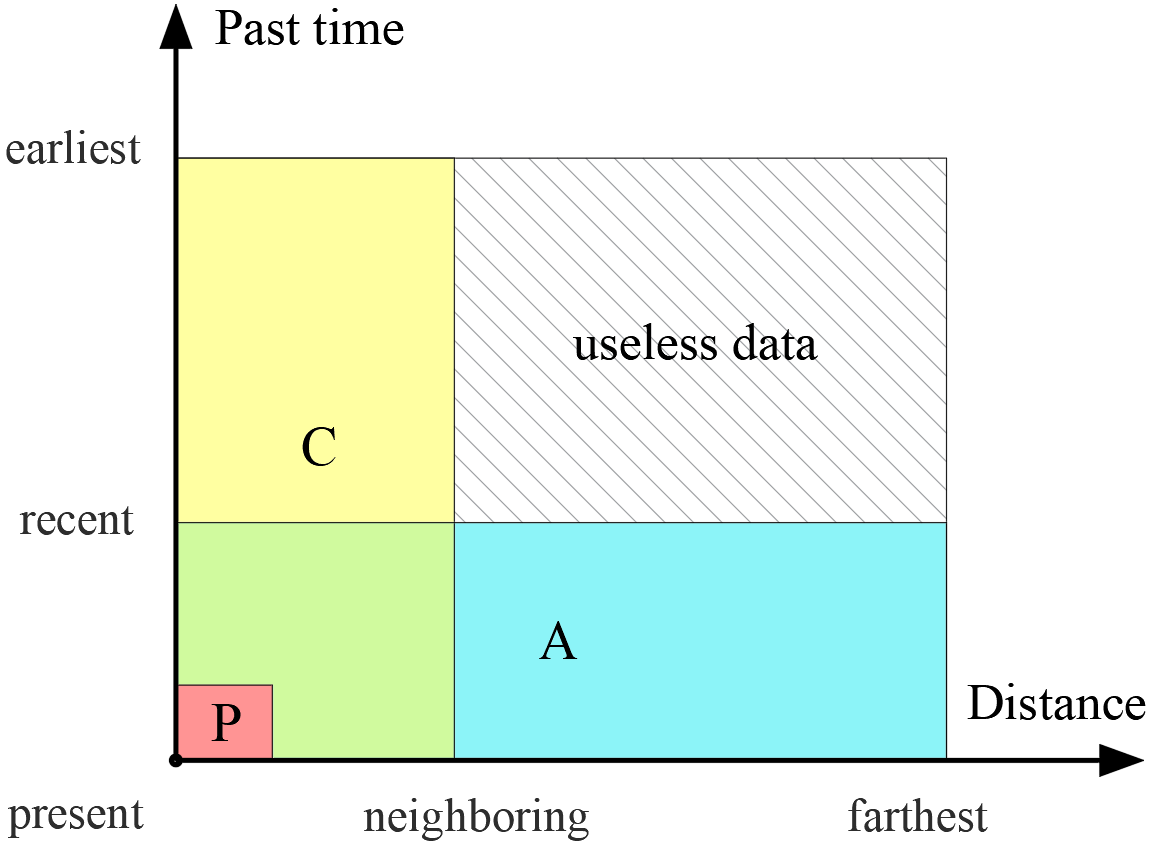}
	\caption{Data classification when matching present probe.}
	\label{F:data}
\end{figure}

Considering different information extraction methods, this paper classified all data into four groups, and their layout are sketchily shown in Fig. \ref{F:data}: %sorts all data into four layers (COV from high to low): 
1) vehicle's \underline{P}resent  data (P-data), 2) \underline{C}ollaborative data (C-data), including vehicle's own historical data and neighboring vehicles' historical data, 3) \underline{A}ll vehicles' recent data (A-data), and 4) other useless data (unused data when matching a trajectory). 
 Note that there exist some overlapping areas, and the data's ``belonging'' is not a fixed characteristic. %We use different methods to extract information from the data, and their details are in the Section \ref{S:SM}. 
A piece of data can be ``useless'' to one trajectory matching, but ``useful'' to another trajectory matching. 
Different from most existing studies that make use of limited information to match trajectories, this paper explores a framework to take full advantage of ``big'' data. On one hand, to enhance the accuracy of MM, we design a corresponding ``scoring'' method for each useful data group, and then fuse the scores to infer the trajectory. On the other hand, to improve the searching efficiency, we adopt a modified top-K shortest-path algorithm within an ellipse region to filtrate the candidate paths. 

%This paper proposes a new MM algorithm to make full use of known data (GNSS probes and prediction results). When matching a trajectory, we firstly sort all data into four groups based on data's COV. Second, we design corresponding methods to extract information (scores for different matching probabilities) from useful data with different COVs. Inspired by the information fusion method \cite{lu2020information}, we also design an adaptive learning method to fuse the extracted information. We then  

The rest of this paper is organized as follows: the next section describes our problem. The third section designs the MM score methods. %and then designs the weights$\backslash$coefficients' update or calibration methods;
The fourth section infers the path and matches the probe. The fifth section uses real-world data to calibrate and test our method. The last section concludes the paper.

\section{Problem Description}

\begin{figure}[h!]
	\centering
	\includegraphics[width=0.55\linewidth]{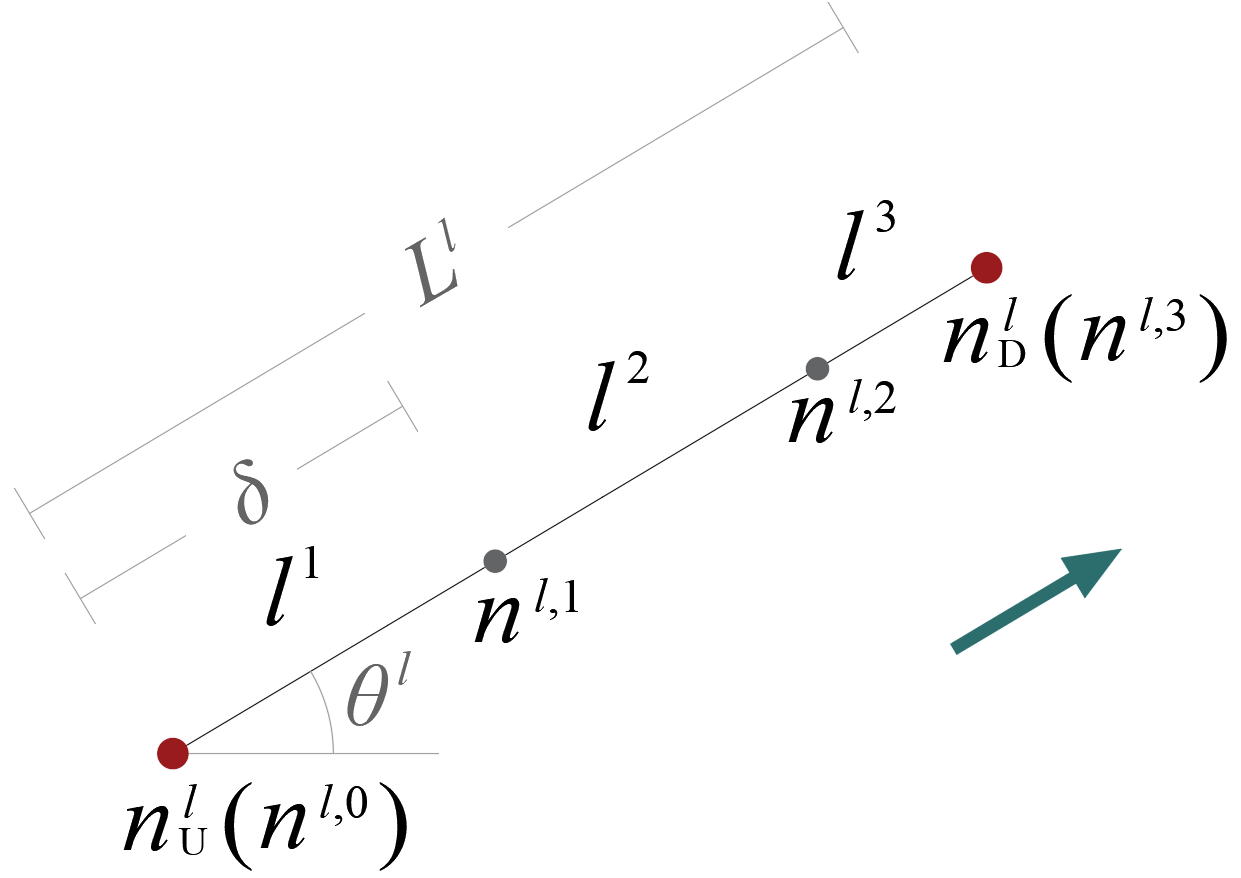}
	\caption{An example of link $l$ consisting of edges $l^1$, $l^2$ and $l^3$.}
	\label{F:link}
\end{figure}

%\subsection{Network description}
Consider an urban traffic network represented by the directed graph $\GX = (\NX,\LX)$, where $\NX$ is set of intersection nodes  and $\LX$ is set of links. %$\GX = (\SX,\A)$. $\SX$ is a set of directed road segments, and any two-way road contains two segments with opposite directions. %Given $r,s \in \SX$, 
%$\A = [a_{sr}]_{|\SX| \times |\SX|}$ is an adjacent matrix of segments, where $a_{sr} =1$ if segment $r$ is an adjacent downstream road of segment $s$; and $0$ otherwise. 
Each $l \in \LX$ has an attribute tuple $\ring{A}(l) = \big(L^l, \theta^l, n^l_\mathrm{U}, n^l_\mathrm{D}\big)$, where $L^l$ is its length, $\theta^l$ is its direction angle (between link and east), $n^l_\mathrm{U} \in \NX$ is its upstream intersection node, and $n^l_\mathrm{D} \in \NX$ is its downstream intersection node.
%To improve MM efficiency, we further split 
Any link $l \in \LX$ is further split into $M%(s) 
$
continuous sub-links (named \emph{edges}): $l^1,l^2,\cdots,l^M$ (Fig. \ref{F:link} shows an example with $M=3$). The node between $l^e$ and $l^{e+1}$ is link's node, denoted by $n^{l,e}$. %\red{$s_1,s_2,\cdots,s_{M}$ (Fig. \ref{F:segment} shows an example with $M=3$).We denote %the start-node of $s$ as $n^s_0$, and 
%the node between $s_e$ and $s_{e+1}$ as $n^s_e$. Hence $s$ contains $M+1$ nodes: $n^s_0, n^s_1, \cdots, n^s_M$. We have $n^s_M = n^s_\mathrm{T} = n^r_0$ if $a_{sr}=1$.} 
Assume the split length is a constant, $\delta$, then $M = \lceil {L^l}/{\delta}\rceil$, %edge lengths $L^s_1 = L^s_2 = \cdots = L^s_{M-1} =\delta$, and edge length 
%$L^s_M = L^s - (M-1) * \delta$. 
edge lengths $L^{l,1} = L^{l,2} = \cdots = L^{l,M{-}1} =\delta$, and edge length $L^{l,M} = L^l - (M{-}1) {*} \delta$. 
\vspace*{-0.3cm}

% 实验部分，描述如何数据处理，剔除停滞点。
Given a probing interval (PI), denoted by $\Delta t$, % \red{GNSS} receiver can get a sequence of corresponding \red{GNSS} probes,
a trajectory $\J$ can be described by a sequence of GNSS probes $\bm{p}^\J = [p_0, p_1, \cdots]$. %, which indicates a vehicle's trajectory. 
Each $p_i$ has an attribute tuple $\ring{A}(p_i) = \big(t^{p_i}, v^{p_i}, \theta^{p_i},x^{p_i},y^{p_i} \big)$, where %$\J$ represents a  trajectory, 
$t$ is its timestamp, $v$ is its speed, $\theta$ is its bearing angle, and $x$ and $y$ represent its longitude and latitude information.
Besides, each $\J$ has an attribute tuple $\ring{A}(\J) = \big(\VV^\J, p_0^\J, p_\mathrm{E}^\J, t_0^\J, t_\mathrm{E}^\J, \bm{p}^\J \big)$, where $\VV$ is vehicle id, $p_0$ (or $p_\mathrm{E}$) is the start-point (or end-point), and $t_0$ (or $t_\mathrm{E}$) is the start-time (or end-time). If a trajectory is unfinished, $p_\mathrm{E}$, $t_\mathrm{E}$ and $\bm{p}$ will keep updating. Our MM algorithm aims to project the probes of any trajectory to specific edges, and infer the path in between adjacent probes.

\begin{figure}[h!]
	\centering
	\includegraphics[width=0.9\linewidth]{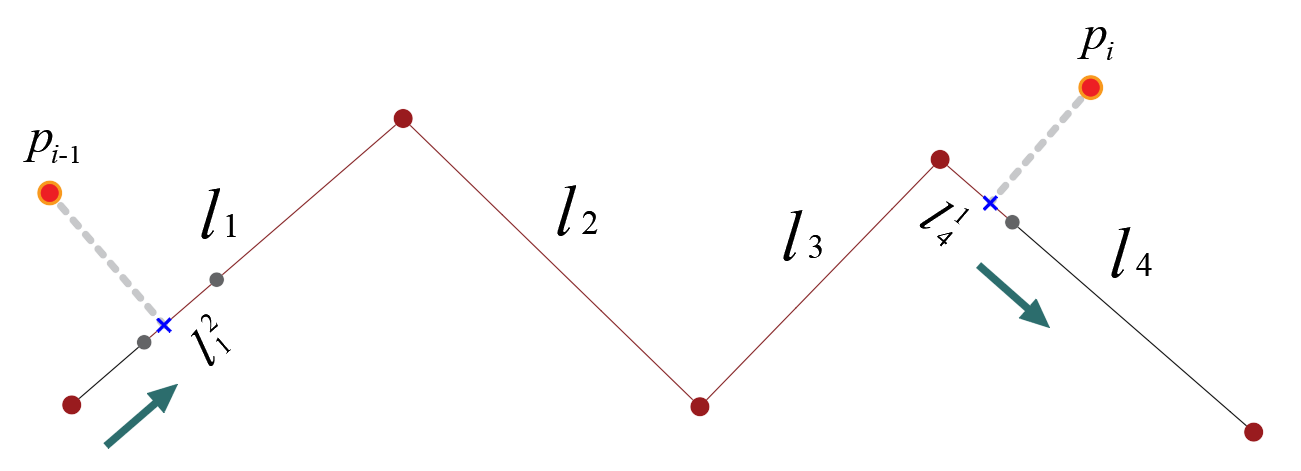}
	\caption{Example of a path $\PX = [l_1^2,l_2,l_3,l_4^1]$ (from $p_{i-1}$ to $p_i$).}
	\label{F:path}
\end{figure}

Suppose GNSS probe $p_{i-1}$ is projected to candidate start-edge %$s(1)_{e}$
$l_1^{e_{i-1}}$, GNSS probe $p_i$ is projected to a candidate end-edge %$s(J)_{f}$
$l_I^{e_i}$, and a candidate path, $\PX$, passes though links %$s(1), s(2), \cdots, s(J)$
$l_1, l_2, \cdots, l_I$, we can express $\PX$ by %a sequence of passing nodes: $\PX = [n^{s(1)}_{e},n^{s(2)}_0, n^{s(3)}_0,\cdots,n^{s(J)}_0,n^{s(J)}_{f-1}]$. 
its start-edge, passed links and end-edge: $\PX = [l_1^{e_{i-1}},l_2, l_3, ...,l_{I-1}, l_I^{e_i}]$.    
Fig. \ref{F:path} shows an example with $I=4$ ($\PX$ passes 4 links). Note that the length of path $\PX$ (denoted by $L^{\PX}$) is the %sum of 1) the passed edges$\backslash$segments from $p_{i-1}$'s projection %node $[n^{s(1)}_{e}$ to node $n^{s(2)}_0$, to $n^{s(3)}_0$, $\codts$,and then to node $n^{s(J)}_{f-1}$, 2) the length 
passed distance from start-projection to end-projection.
%\vspace*{-0.4cm}
% 图修改：p_{i-1} 和 p_i；路径长度示意, p 挪动一些位置。 J 改成 I

\begin{figure}[h!]
	\centering
	\includegraphics[width=0.98\linewidth]{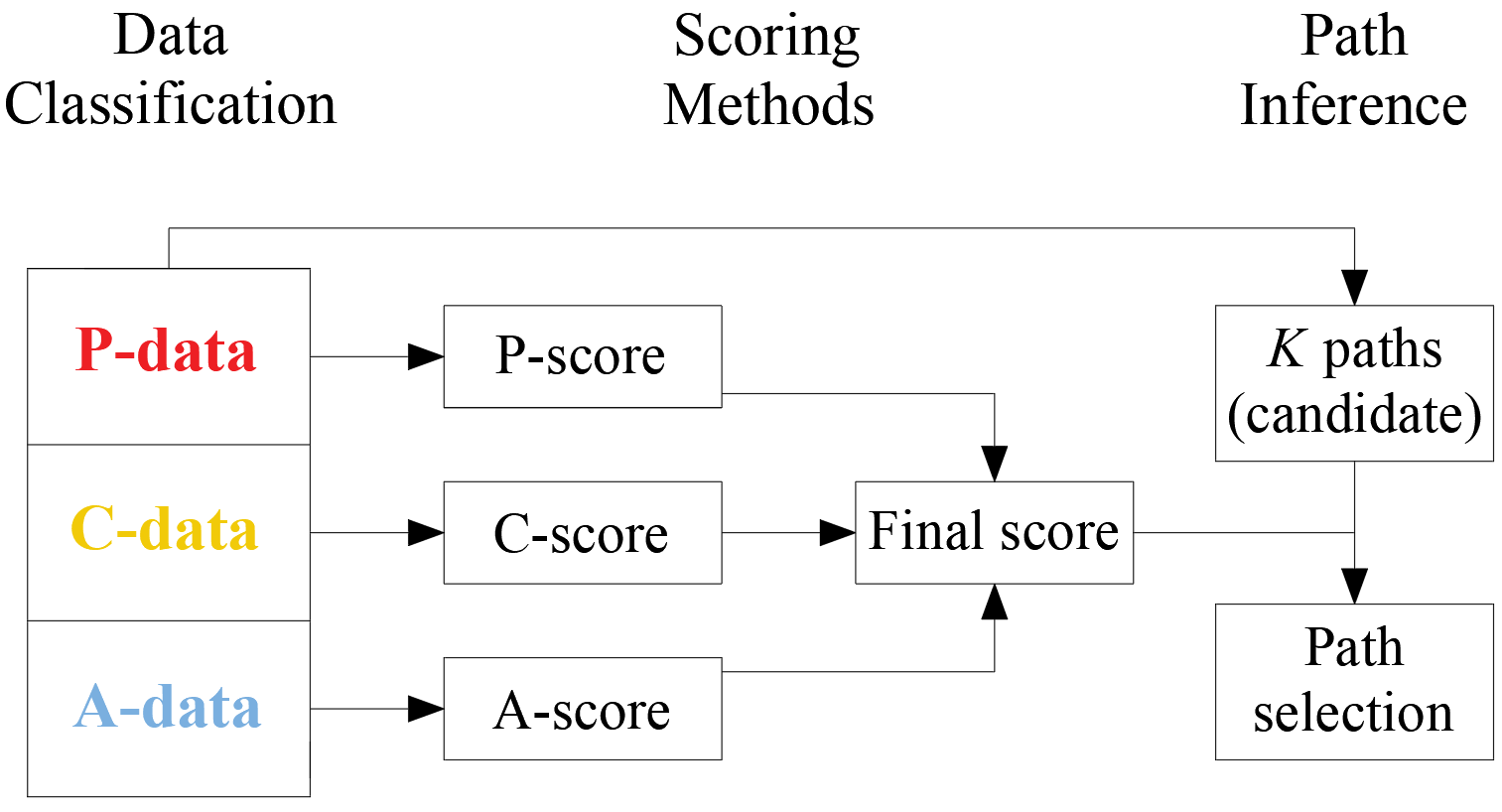}
	\caption{Framework of proposed method.}
	\label{F:score}
\end{figure}

Our method combines path inference and probe projection together. When projecting a probe $p_i$ of trajectory $\J$, we first use the shortest-path method to choose $K$ number of candidate paths (from probe $p_{i-1}$ to probe $p_i$) whose set is $\bm{\PX}_\J^i$ (paths could have different projected start- or end-edges). Then we score all candidate paths and finally pick the one with the highest score as the inferred path. The projected point (edge) of probe $p_i$ is then determined by the inferred path. The framework of our method is shown in Fig. \ref{F:score}. To guarantee the ``fairness'' of 
our scoring method, there are three ``judges'' that come from different data groups and score all candidate paths, respectively. The weighted mean score of a path is its final score. We will show the details in next section.
 %提及 scores 取值从0到1
% 提一下， p_{i-1} 和 p_i 不是重合点，否者跳过该步骤，并让 i = i + 1。

The main parameters in this paper are summarized in Table \ref{t:notation}. 
%\subsection{Notation summary unfinished}
\begin{table}[h!]
	%\begin{mdframed}[hidealllines=true,backgroundcolor=yellow]
	\caption{Notations in this paper}
	\label{T:Notation}  %of Chapter \ref{chap:lane} \label{lane_t1}
	\begin{tabular} {c p{0.69\linewidth}} 
		\hline
		\\[-0.8em]
		\textbf{Parameter} & \textbf{Description} \\
		\\[-0.9em]
		\hline
		\\[-0.9em]
		$\ring{A}$ & attribute function \\
		$\bm{A}$, $A$ & adjacent matrix \\
		% $a$ & one connecting relationship \\
		$e$ & an edge's order in a link \\
 		$\GX$ & graph \\
 		$G$ & a collaborative group \\
 		$\mathcal{H}$, $\overline{\mathcal{H}}$ & historical usage frequency of an edge in MM results, weighted historical usage frequency \\
 		$h_{\max}$, $h_{\min}$ & max value or min value of  weighted historical usage frequency of an edge \\
 		$I$ & passed links \\
 		$i$ & order of a probe in a trajectory \\
 		$\J$ & a trajectory \\
 		$K$ & number of candidate paths \\
 		$k_{\mathrm{max}}$ & max time step in traffic distribution prediction \\
		$\LX$ & set of directed links \\
		$\bm{L}$ & Laplacian matrix \\
		$L$ & length \\
		$l,m$ & a link \\
		$M$ & number of a link's sub-links (edges) \\
		$\NX$ & set of intersection nodes \\
		$N^\mathrm{S}$,$N^\mathrm{E}$ & from-node or end-node of a candidate edge set\\
		$n$ & node \\ 
		$\PX$, $\PX^*$, $\bm{\PX}$ & a candidate path, a real path, and a candidate path set\\
		$\overline{P}$ & mean vehicle percentage of a path \\
		$p$, $\bm{p}$ & probe or probe sequence \\
		$R$ & radius of a probe's vicinity area \\
		$r_\mathrm{s}$,$r_\mathrm{t}$ & spatial radius or temporal radius in collaborative group \\ 
		$S_\mathrm{P}$, $S_\mathrm{C}$, $S_\mathrm{A}$ & values of P-score, C-score or A-score\\
		$t$ & time \\
		$\Delta t$ & probing interval\\
		$\Delta T$ & a short period of time \\
		$v$ & speed \\
		$W_{\mathrm{speed}}$, $W_{\mathrm{bear}}$ & speed weight, bearing weight\\
		$W_\mathrm{P}$, $W_\mathrm{C}$, $W_\mathrm{A}$ & weights of P-score, C-score or A-score\\
		$w_\mathrm{c}$ & weight coefficient in collaborative group \\
		$X$, $\hat{X}$ & percentage of vehicles allocated at link, or its predicted value \\
		$x$ & longitude information \\ 
		$Y$,$\hat{Y}$ & accuracy, estimated accuracy \\
		$y$ & latitude information \\
		$\lambda$ & speed coefficient\\
		$\delta$ & split length \\
		$\mathcal{E}$, $\mathcal{E}'$ & mean square error\\ 
		$\theta$ & direction angle\\
		$\Delta \tau$ & traffic state updating interval \\
		\\[-0.9em]				
		\hline
	\end{tabular}
	\label{t:notation}
	%\end{mdframed}
\end{table}

\section{Scoring Methods}
\label{S:SM}

\subsection{P-score}

The first ``judge'', P-score, denoted by $S_\mathrm{P}$, comes from P-data, which consists of last probe ($p_{i-1}$) and this probe ($p_i$). Its equation includes two parts: speed weight, denoted by $ W_\mathrm{speed}$, and bearing weight, denoted by $W_\mathrm{bear}$. For a path $\PX \in \bm{\PX}_\J^i$, its P-score is 
\begin{equation}
	S_\mathrm{P}^\PX = W_\mathrm{speed}^\PX W_\mathrm{bear}^\PX \times 100\%.
	\label{E:OP-score}
\end{equation}

%\textit{a)} 
\subsubsection{Speed weight} The mean speed from $p_{i-1}$ to $p_i$ is estimated by average speed at both probes. %$(v^{p_{i-1}}+v^{p_i})/2$. 
For a given candidate path $\PX$, its mean speed equals to the traveling distance $L^\PX$ over $\Delta t$. Hence, their speed difference %$\Delta v$ is estimated by:
is $|(v^{p_{i-1}}+v^{p_i})/{2} - {L^\PX}/{\Delta t}\big|$. Inspired by \cite{song2018hidden} which indicated that the distribution of speed difference is negative exponential, the speed weight  $W_\mathrm{speed}^\PX$ is designed as:
\begin{equation}
	W_\mathrm{speed}^\PX = \exp\big( -\lambda \cdot \Big|\frac{v^{p_{i-1}}+v^{p_i}}{2} - \frac{L^\PX}{\Delta t}\Big|\big),
	\label{E:speedweight}
\end{equation}
where $\lambda$ is a positive coefficient. When speed difference is 0, 	$W_\mathrm{speed}^\PX = 1$; when speed difference is large, $W_\mathrm{speed} ^\PX \approx 0 $.
%The speed score is maximized when $	\Delta v =0$. 
% in simulation. Discuss how to estimate $\Lambda$.

%\textit{b)} 
\subsubsection{Bearing weight} We take cosine function to estimate $\PX$'s bearing weight $W_\mathrm{bear}^\PX$. Suppose candidate end-edge of $\PX$ (from $p_{i-1}$ to $p_i$) locates at link $l$,
\begin{equation}
	W_\mathrm{bear}^\PX = \max\big\{\cos \big( |\theta^{p_i}-\theta^l| \big) , 0 \big\},
	\label{E:bearingweight}
\end{equation}
where $\theta^{p_i}$ is $p_i$'s bearing and $\theta^l$ is the direction of $l$. When when $p_i$ and $l$'s inclination is 0, 	$W_\mathrm{bear}^\PX$ is maximized and equal to 1; when their inclination is equal to or larger than $90^\circ$, $W_\mathrm{bear}^\PX = 0$. %Vehicle's bearing angle should be similar to segment's direction of its real road direction.  

\subsection{C-score}

The second ``judge'', C-score, denoted by $S_\mathrm{C}$, comes from C-data, which consists of two parts: ego vehicle's own historical MM results and neighbors' historical MM results. 

A vehicle's historical data can reflect its traveling habit. If a path has been chosen several times in the past, it is likely to be chosen again in the future trip. Furthermore, other vehicles' trips with similar OD and traveling period, usually have the same route choices, %their routes are usually highly overlapping because the best choices are usually the same.
and their historical MM results can also contribute to ego vehicle's MM. %bian2020trajectory
Therefore, we define $\J$'s \emph{collaborative group} $G_\J$ that consists of itself and all its neighboring trajectories, $G_\J = \big\{\J'{:}~||p_0^\J{-}p_0^{\J'}{||} {\leq} r_\mathrm{s}, ||p_\mathrm{E}^\J{-}p_\mathrm{E}^{\J'}{||}{\leq} r_\mathrm{s}, |t_0^\J{-}t_0^{\J'}{|} {\leq} r_\mathrm{t}, |t_\mathrm{E}^\J{-}t_\mathrm{E}^{\J'}{|} \\ {\leq} r_\mathrm{t}\big\}$ where $r_\mathrm{s}$ ($r_\mathrm{t}$) is spatial (temporal) radius. %We will use the historical  MM results from vehicle's own and collaborative data to improve MM accuracy.

Suppose $\PX$'s historical usage frequency is $\overline{\cH}_\J^{\PX}$. Its C-score is %there are $K$ number of candidate paths from $p_{i-1}$ to $p_i$ for $\J$, whose set is denoted by $\bm{\PX}_\J^i$. For each $\PX \in \bm{\PX}_\J^i$, its CH-score, denoted by $S_\mathrm{OH}^\PX$, is calculated by
\begin{equation}
	S_\mathrm{C}^\PX = 
	\left\{
	\def\arraystretch{1.35}
	\begin{array}{cl}
		\frac{\overline{\cH}_\J^{\PX} - h_{\min}}{ h_{\max} - h_{\min} } \times 100\% & ~ \mathrm{if} ~ h_{\max} > h_{\min} \\
		0 &  ~\mathrm{otherwise}
	\end{array},
	\right.
	\label{E:C-score}
\end{equation}
where $h_{\min} = \min\big\{ \overline{\cH}_\J^{\PX’}{:}~{\PX'\in\bm{\PX}_\J^i} \big\}$, and $h_{\max} = \max\big\{ \overline{\cH}_\J^{\PX’}{:} $ $ \PX' \in \bm{\PX}_\J^i \big\}$.
Clearly, $S_\mathrm{C}^\PX$ is positively correlated with $\overline{\cH}_\J^{\PX}$.

\textit{Estimation of historical usage frequency:} Suppose at time $t_0^\J$, we aggregate $\J$'s historical MM results of ego vehicle and record the frequency it has passed edge $l^e$, denoted by $\cH_\J^{l,e}$. Similarly, we also record $\cH_{\J'}^{l,e}$ for $\J' \in G_\J$. 
We denote the number of edges that $\PX$ passes by $|\PX|_\mathrm{edge}$, and the weighted historical usage frequency for $\PX$ (denoted by $\overline{\cH}_\J^{\PX}$) is estimated by 
\begin{equation}
	\overline{\cH}_\J^{\PX} = \frac{\sum_{l^e \in \PX} \cH_\J^{l,e} + w_\mathrm{c}\sum_{\J' \in G_\J\backslash\J} \sum_{l^e \in \PX} \cH_{\J'}^{l,e}}{(1+w_\mathrm{c}|G_\J-1|)\cdot|\PX|_\mathrm{edge}},
	\label{E:hf}
\end{equation}
where $w_\mathrm{c} \in [0,1]$ is a weight coefficient for the neighbor trajectories. $w_\mathrm{c} = 0$ implies that we only consider the ego vehicle's own historical trajectory, and $w_\mathrm{c} = 1$ implies we put same significance to both ego vehicle and other vehicles with neighboring trajectories.

\subsection{A-score}

The third ``judge'', A-score, denoted by $S_\mathrm{A}$, comes from A-data, which consists of all MM results in recent $\Delta T$ time. 

If we know that link $l$ would have lots of distributed traffic (compared with its neighboring links) during time $\Delta \tau$, we can predict that a path passing $l$ is more likely to be the true path during $\Delta \tau$. In other words, the results of traffic state prediction can contribute to the MM's accuracy. 
Suppose the prediction of traffic distribution updates every interval $\Delta \tau$. We denote $\hat{X}^l$ as the percentage of vehicles allocated at link $l$ over all vehicles in prediction. The mean vehicle percentage of path $\PX$, denoted by $\overline{P}^\PX$, is estimated by:
\begin{equation}
	\overline{P}^\PX = \frac{\sum_{l \in \PX} \hat{X}^l}{|\PX|_\mathrm{link}},
	\label{E:P}
\end{equation}
where $|\PX|_\mathrm{link}$ is the number of links $\PX$ passes. The A-score for $\PX$ is 
\begin{equation}
	S_\mathrm{A}^\PX = 
	\left\{
	\def\arraystretch{1.35}
	\begin{array}{cl}
		\frac{\overline{P}^\PX - P_{\min}}{ P_{\max} - P_{\min} } \times 100\% & ~ \mathrm{if} ~ P_{\max} > P_{\min} \\
		0 &  ~\mathrm{otherwise}
	\end{array},
	\right.
	\label{E:A-score}
\end{equation}
where  $P_{\min} = \min\big\{ \overline{P}^{\PX'}{:}~{\PX'\in\bm{\PX}_\J^i} \big\}$, and $P_{\max} = \max\big\{\overline{P}^{\PX'}{:} $ $ \PX' \in \bm{\PX}_\J^i \big\}$. Clearly, $S_\mathrm{A}^\PX$ is positively correlated with $\overline{P}^\PX$.

We offer two methods to predict $\hat{X}$.
\subsubsection{Prediction method without NN}

Suppose at time $j\Delta\tau $ ($j {\in} \NN^+$), we aggregate all locations in MM results during interval $\big[(j{-}1)\Delta\tau, j\Delta\tau\big)$ and record vehicles' percentage vector $X_j {=} \big(X_j^1, X_j^2,\cdots, X_j^{|\LX|} \big)^\mathrm{T}$, where $\sum_{l \in \LX} X_j^l = 1$. Note that, we set every link to have one vehicle before aggregation, hence $X_j^l>0$.

If $X_{j-1}$ is given but $X_j$ is unknown, a simple way to predict $X_j$ is using weighted mean value of historical $X$,
\begin{equation}
	\hat{X}_j = \sum_{k = 1}^{k_{\mathrm{\max}}}\gamma_k X_{j-k},
	\label{E:hatX}
\end{equation}
where $\hat{X_j}$ is the prediction vector, $k_{\mathrm{\max}}$ is the max considered time steps with $k_{\mathrm{\max}} = \min\big\{\lceil \Delta T/ \Delta \tau \rceil,j\big\}$, and $\gamma$ is a temporal decay ratio with $\sum_{k = 1}^{k_{\mathrm{\max}}}\gamma_k = 1$. However, this prediction method may have a high error and ignores the graph information. Hence, we also offer a graph-based prediction method using NN. 

\subsubsection{Prediction method with NN}
Similar to \cite{defferrard2016convolutional,cui2020graph}, we define a symmetric adjacent matrix $\bm{A} = [A_{l,m}]_{|\LX| \times |\LX|}$, where 
$A_{l,m} = 1$ if link $l$ and $m$ share a common intersection node; otherwise, $A_{l,m} = 0$.  %Note that undirected $\bm{A}$ does not influence the set of candidate paths because paths is searched in the directed graph $\GX$. 
We also define the Laplacian matrix as $\bm{L} = \mathrm{diag}\big( \sum_\iota \A_{\small~\bigcdot,\iota}\big)-\bm{A} $. $\bm{L}$ can be diagonalized as $\bm{L} = U\Lambda U^\mathrm{T}$ by its diagonal eigenvalue matrix $\Lambda = \mathrm{diag}(\lambda_1,\lambda_2,\cdots,\lambda_{|\LX|})$ and its eigenvector matrix $U = [u_1,u_2,\cdots,u_{|\LX|}]$ satisfying $\bm{L}u_\iota = \lambda_\iota u_\iota$. This paper adopts spectral graph Markov NN (SGMN) considering $k_{\mathrm{\max}}$ time steps \cite{cui2020graph}, and we have
\begin{equation}
	\hat{X}_j = \sum_{k = 1}^{k_{\mathrm{\max}}}\gamma_k U \bm{\Lambda}_k U^\mathrm{T}X_{j-k},
	\label{E:hatX2}
\end{equation}
where $\bm{\Lambda}_k$ is a learnable diagonal matrix with a initial value equal to $(\Lambda)^k$ (which means k power of $\Lambda$), and any $U \bm{\Lambda}_k U^\mathrm{T}$ %(with $j{-}k_{\mathrm{\max}}{\le}k {\le} j{-}1$) 
is a hidden layer%(the other parameters are introduced in the prediction method without NN)
. Note that we only have the data of $X_k$ with $k = 0,1,\cdots,j{-}1$. Hence, in the training process, we let $j’ {:}{=}j{-}1$, and $X_{j'}$ or $\hat{X}_{j'}$ is the true or predicted value. In addition, we choose mean square error %absolute percentage error 
(denoted by $\mathcal{E}$) as the loss function to update $\bm{\Lambda}_k$ in back-propagation step, we have
%\begin{equation}
%	\mathcal{E} = \frac{1}{|\LX|}\sum_{l=1}^{|\LX|}\Big|\frac{X_{j'}^l-\hat{X}_{j'}^l}{X_{j'}^l}\Big|.
%	\label{E:Error}
%\end{equation}
\begin{equation}
	\mathcal{E} = \frac{\sum_{l=1}^{|\LX|}\big(X_{j'}^l-\hat{X}_{j'}^l\big)^2}{|\LX|}.
	\label{E:Error}
\end{equation}

%MM results are useful to evaluate traffic distribution and further predict the evolution of future distribution. In probability, a path with more predicted traffic will also be more likely to be chosen by current MM vehicle. Hence, we can utilize this predicted traffic information to improve MM accuracy.  

\begin{comment}
Suppose from time $j\Delta\tau$ to $(j+1)\Delta\tau$, there are $H_j^s$ number of vehicles allocated at segment $s$ with MM, where $\Delta \tau$ is aggregation interval and $j \in \NN_0$. We use the softmax function to initialize the allocation probability (denoted by $X$),  
\begin{equation}
	X_j^s = \frac{\exp(c_jH_j^s)}{\sum_{s\in \SX} \exp(c_jH_j^s) },
	\label{E:X}
\end{equation}
where coefficient $c_j =  1/\max\big\{\underset{s \in \SX}{\max} H_j^s  - \underset{s \in \SX}{\min} H_j^s,1\big\}$. We define the probability vector (of $j$th interval) by $\bm{X}_j = \big[X_j^1, X_j^2, \cdots, X_j^{|\SX|} \big]$. We further define a symmetric adjacent matrix $\bm{A} = [a_{rs}]_{|\SX| \times |\SX|}$. 
$a_{rs} = 1$ if segments $r$ and $s$ share a common intersection node; otherwise, $a_{rs} = 0$. Note that undirected $\bm{A}$ does not influence the set of candidate paths because paths is searched in the directed graph $\GX$. 

% 把 GMN 放到 simulation
\end{comment}

\subsection{Final-score}
Given three ``judge'' scores $S^\PX_\mathrm{P}$, $S^\PX_\mathrm{C}$, $S^\PX_\mathrm{A}$, we denote the final score function as $\F(\PX)$, 
\begin{equation}
	\F(\PX) = W_\mathrm{P}S^\PX_\mathrm{P}+W_\mathrm{C}S^\PX_\mathrm{C}+W_\mathrm{A}S^\PX_\mathrm{A},
	\label{E:F}
\end{equation}
where $W_\mathrm{P}$, $W_\mathrm{C}$ and $W_\mathrm{A}$ are corresponding weight satisfying $W_\mathrm{P}+W_\mathrm{C}+W_\mathrm{A} = 1$. 

We list two methods to set weights as follows.
\subsubsection{Equal weights} 
The simplest way is to assign the same value to all weights: $W_\mathrm{P}=W_\mathrm{C}=W_\mathrm{A} = 1/3$.
 
%\subsubsection{Gradient weight} Another way is to assign gradient weights to different ``judges''. A ``judge'' that is more trustworthy should have higher weight. In our paper, we have calibrated them that is demonstrated to perform well: $[W_\mathrm{P},W_\mathrm{C},W_\mathrm{A}]=[0.3,0.5,0.2]$.

\subsubsection{Calibrated weights}
%Both methods above ignore the difference in accuracy among raw data. 
If  ``Big'' data consists of low-frequency data and high-accuracy \& high-frequency data, we can regard the high quality data as the ``anchor'' to calibrate the weights with machine learning. It includes three steps:
\begin{enumerate}[label=\arabic*)]
\item Use the shortest-path method (e.g., classical Dijkstra method \cite{dijkstra1959note}) to get the ``real'' paths (denoted by $\PX^*$) for the high-quality data. Based on the high-quality data, one can artificially create several groups of low-frequency data by removing some probes (with fixed interval). Between probe $p_{i-1}$ and probe $p_i$ in a created low-frequency data, one can obtain at most $K$ number of candidate paths (using top-K shortest loopless paths introduced in next section, which is similar to \cite{yen1971finding}).

\item For each path $\PX$, its accuracy, denoted by $Y^\PX \in [0,1]$, is estimated by
\begin{equation}
	Y^\PX = \frac{\sum_{l^e \in \PX}\mathds{1}\big( l^e \in \PX^* \big)}{\big| \PX \big|_{\mathrm{edge}}}
	\label{E:Y}
\end{equation}
where $l^e$ is $\PX$'s edge, $\mathds{1}( {\cdot} )$ is an indicator function equal to 1 if true and 0 otherwise.

\item Based on the scoring methods, we can get score vector $S^\PX = [S^\PX_\mathrm{P},S^\PX_\mathrm{C},S^\PX_\mathrm{A}]$ for path $\PX$. One can build a simple NN structure with a fully connected layer to estimate the weight for each score, with%and choose ReLU as the activation function, 
\begin{equation}
	%\hat{Y}^\PX = \max\big\{  W (S^\PX)^\mathrm{T} + B, 0 \big\},
	\hat{Y}^\PX = W (S^\PX)^\mathrm{T} + B,
	\label{E:hatY}
\end{equation}
where $\hat{Y}^\PX$ is the predicted accuracy of $\PX$, $(\cdot)^\mathrm{T}$ is transpose of $\cdot$~, $W$ is a $1\times3$ training vector ($[W_\mathrm{P},W_\mathrm{C},W_\mathrm{A}] = W$), and $B$ is the bias. We pick the mean square error (denoted by $\mathcal{E}'$) as the loss function,
\begin{equation}
	\mathcal{E}' = \frac{\sum_{\PX \in \bm{\PX}}(Y^\PX - \hat{Y}^\PX)^2}{|\bm{\PX}|},
	\label{E:error2}
\end{equation}
where $\bm{\PX}$ is the set of candidate paths, and $|\bm{\PX}|$ is the size of path set.

\end{enumerate}

\section{Path inference}
\label{S:PI}

\subsection{Candidate edges}

When matching the path from $p_{i-1}$ to $p_i$, there are two groups of candidate edges: candidate %start-nodes 
edges of $p_{i-1}$, denoted by $\big\{l^e\big\}_{i-1}$, and candidate %end-nodes 
edges of $p_i$, denoted by $\big\{l^e\big\}_{i}$. Usually, $p_{i-1}$ only has one candidate edge %$\big|\big\{l^e\big\}_{i-1}\big| = 1$ 
because last MM iteration has selected the best one as the ``inferred'' edge. %$p_{i-1}$ only has one candidate edge which is the choice in last iteration. %candidate start-node which has been chosen in last iteration. 
Only when we match the path from $p_0$ to $p_1$ could there be %mutiple 
several candidate %nodes
edges of $p_0$. 

\begin{figure}[h!]
	\centering
	\includegraphics[width=0.8\linewidth]{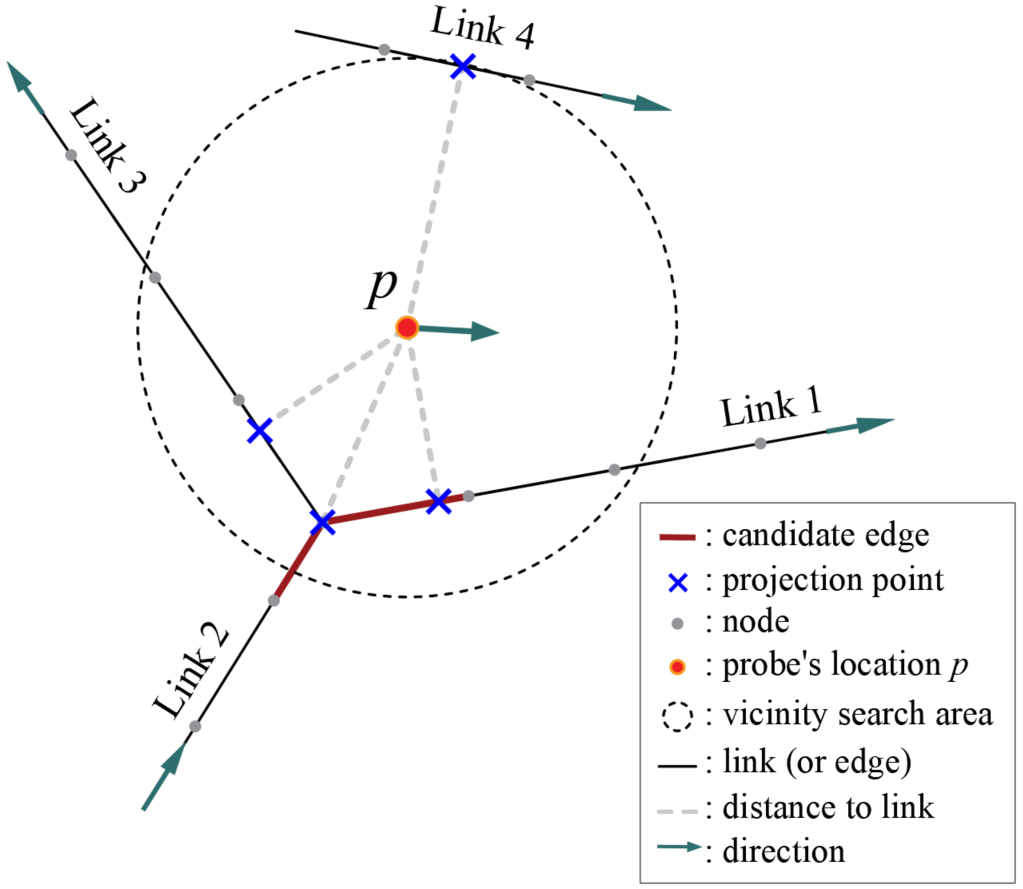}
	\caption{Example of candidate edges in vicinity search area.} %candidate notes
	\label{F:vicinity}
\end{figure}

When our algorithm tries to find probe $p$'s candidate edges, it firstly searches all links in the vicinity of $p$ (radius is $R$). Then, $p$ projects a point to the links. The edge of corresponding projection point is a candidate edge if a) the inclination between the bearing of $p$ and direction of $l$ is smaller than $90^{\circ}$, and 2) one of the edge's two-side nodes is also within $p$'s vicinity area.  
%In this study, a node $n$ (on several segments $\SX^n$) is a candidate node of probe $p$ if 1) it is in the vicinity of $p$ (radius $R$): $||n-p||\leq R$, and 2) there exists a segment $s\in \SX^n$ that makes the inclination between the bearing of $p$ and direction of $s$ smaller than $90^{\circ}$:  $|\theta^p-\theta^s|<90^\circ$ or $270<^\circ|\theta^p-\theta^s|<360^\circ$.
% 描述 如何确定candidate edges % 包括 两个判定
Fig. (\ref{F:vicinity}) shows an example of $p$'s candidate edges. Although there are four projection points, only two of them have candidate edges. There is no candidate edge on link 3 because the inclination between $p$'s bearing and link 3's direction is larger than $90^{\circ}$. There is no candidate edge on link 4 because the corresponding edge's two-side nodes are beyond $p$'s vicinity. %Firstly, we exclude all nodes outside the vicinity area, and there are 6 node left. Secondly, we keep the Segment 1 and 2 and remove Segment 3. Finally, only 4 nodes satisfy two conditions, and they are candidate nodes. 

%椭圆公式，

\subsection{Candidate paths}

During PI $\Delta t$, we assume the ego vehicle travels from $p_{i-1}$ to any point $p$, and then to $p_i$. A conservative assumption is that the trajectories from $p_{i-1}$ to $p$ and from $p$ to $p_i$ are straight-line with max speed. It forms an ellipse region beyond which the vehicle can not reach during $\Delta t$. %Considering real road network, real traveling distance from $p_{i-1}$ to $p_\mathrm{F}$ and from  $p_\mathrm{F}$ to $p_i$, is usually larger than two straight-line distances. We use vehicle's max speed to overestimate its traveling distance and get a feasible region of which the vehicle can not travel out. 
\begin{equation}
	||p - p_{i-1}|| + ||p-p_i|| \leq \max\big\{\max\{v_{i-1},v_i\} \Delta t, 2||p_i - p_{i-1}||\big\}, 
	\label{E:ellipse}
\end{equation}
where$||\cdot||$ is Euclidean distance, $v_i$ is speed at $p_i$, and $p$ is any feasible location. Hence,
Inequality (\ref{E:ellipse}) forms an ellipse region with two focuses $p_{i-1}$ and $p_i$. 
Note that we set a lower-bound of long axis as $2||p_i - p_{i-1}||$ for the ellipse region.

\begin{figure}[h!]
	\centering
	\includegraphics[width=1\linewidth]{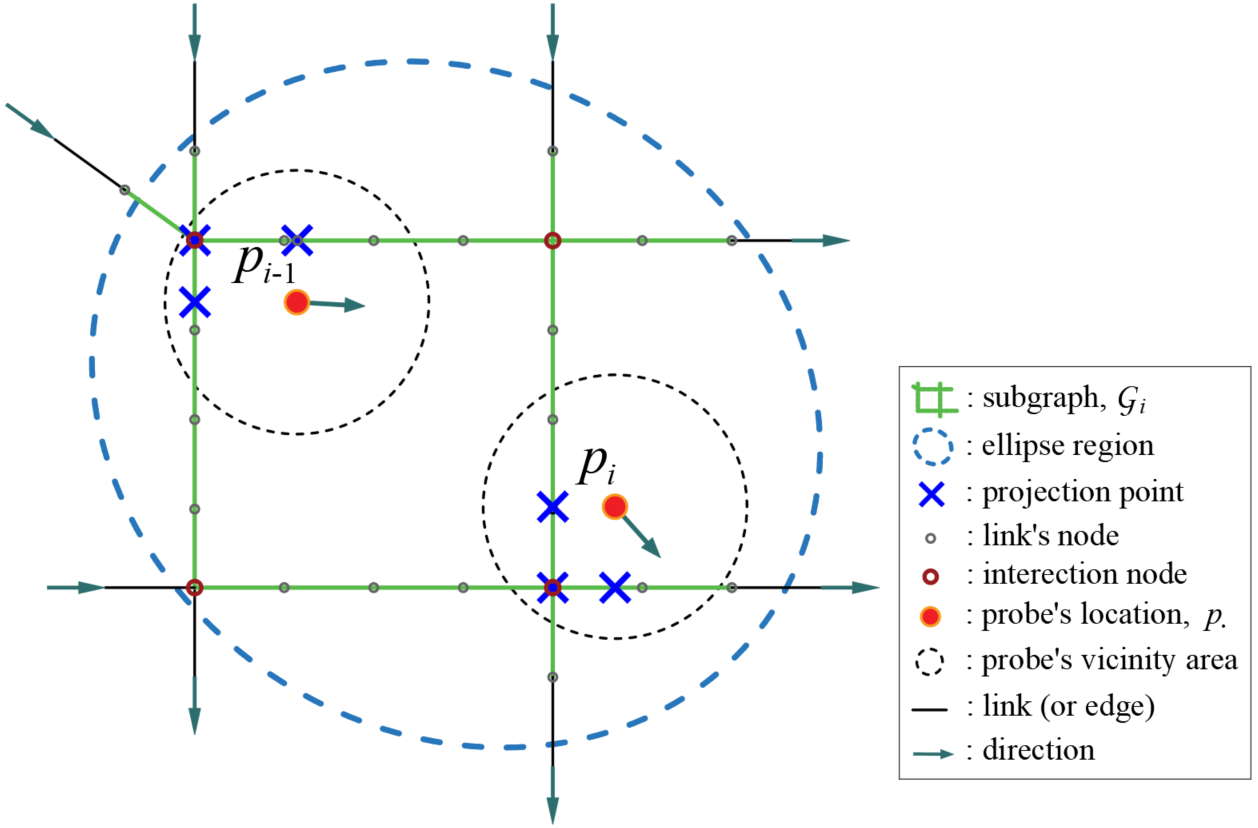}
	\caption{An example of sub-graph network in an ellipse region.}
	\label{F:ellipse}
\end{figure}

We show an ellipse region example in Fig. \ref{F:ellipse}, and trim the graph into a sub-graph, $\GX_i$. 
An edge belongs to %is retained in 
$\GX_i$ if (1) its both-side nodes are within the region (convex), or (2) one of its side nodes is a candidate node of $p_{i-1}$ or $p_i$ and is within the region. All candidate paths should be within $\GX_i$.

%Suppose the Euclidean coordinates of $p_{i-1}$ and $p_i$ in field data are $(x_{i-1},y_{i-1})$ and $(x_{i},y_{i})$, and the coordinate of their middle point is $(x_\mathrm{M},y_\mathrm{M})$. $p_i$ is projected to edge $s^r_e$,  we need to determine the projection of $p_{i+1}$ and the candidate path from $p_i$ to $p_{i+1}$.

\begin{figure}[h!]
	\centering
	\includegraphics[width=0.95\linewidth]{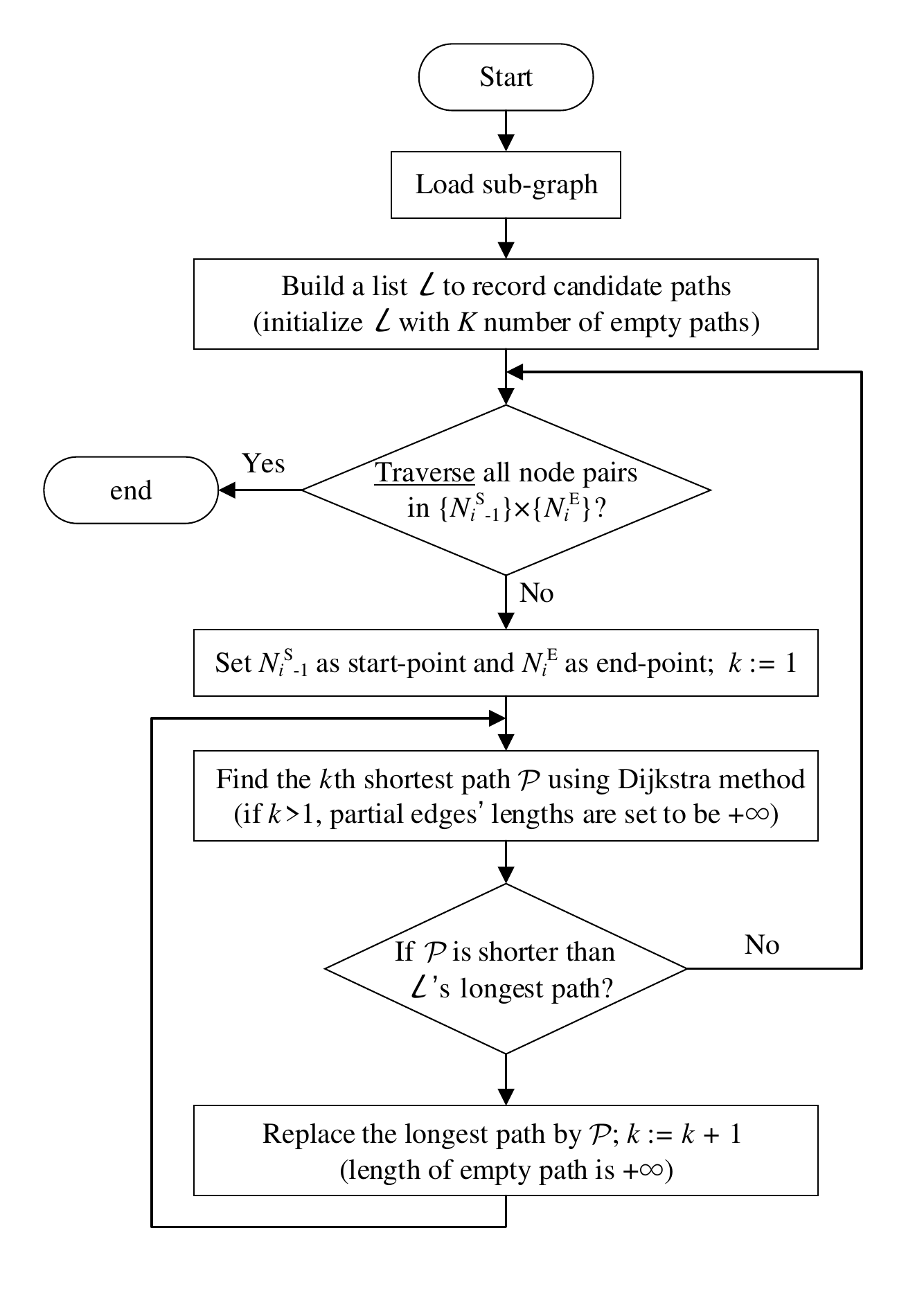}
	\caption{Flowchart of modified top-K shortest-path method.}
	\label{F:shortest_path}
\end{figure}

%all edges of a candidate path should be candidate-path edge (in the region)%from $p_{i-1}$ to $p_i$
Similar to \cite{zheng2012reducing}, we used a modified top-K shortest path method to search paths from $p_{i-1}$ to $p_i$, which is shown in Fig. (\ref{F:shortest_path}). In this flowchart, %we make $\big\{s(e)\big\}_{i-1}$'s head nodes as set of start-points (denoted by $N^s_\mathrm{S}$) and $\big\{s(e)\big\}_i$'s tail nodes as set of end-points (denoted by $N^s_\mathrm{E}$). 
$N_{i-1}^\mathrm{S}$ is a set of $\big\{l^e\big\}_{i-1}$'s from-nodes, and $N_i^\mathrm{E}$ is a set of $\big\{l^e\big\}_i$'s to-nodes.
Except for edge' length, a path's length includes the length from the projection point to node. When the flowchart ends, we save all non-empty paths of the list into the set $\bm{\PX}_\J^i$.
% K 最短路径法确定 candidate paths

\subsection{Path selection}
Combined with Eq. \eqref{E:F}, the path with the highest score is ``inferred'' path (denoted by $\hat{\PX}$) that we select in MM,
%S^\PX_\mathrm{P},S^\PX_\mathrm{C},S^\PX_\mathrm{A}
\begin{equation}
	\hat{\PX} = \underset{\PX \in \bm{\PX}^i_\J}{\arg\max}\F(\PX).
	\label{E:FS}
\end{equation}

\section{Experiment and Evaluation}

\subsection{Data collection and parameter settings}
\begin{figure*}[h!]
	\centering
	\includegraphics[width=0.95\linewidth]{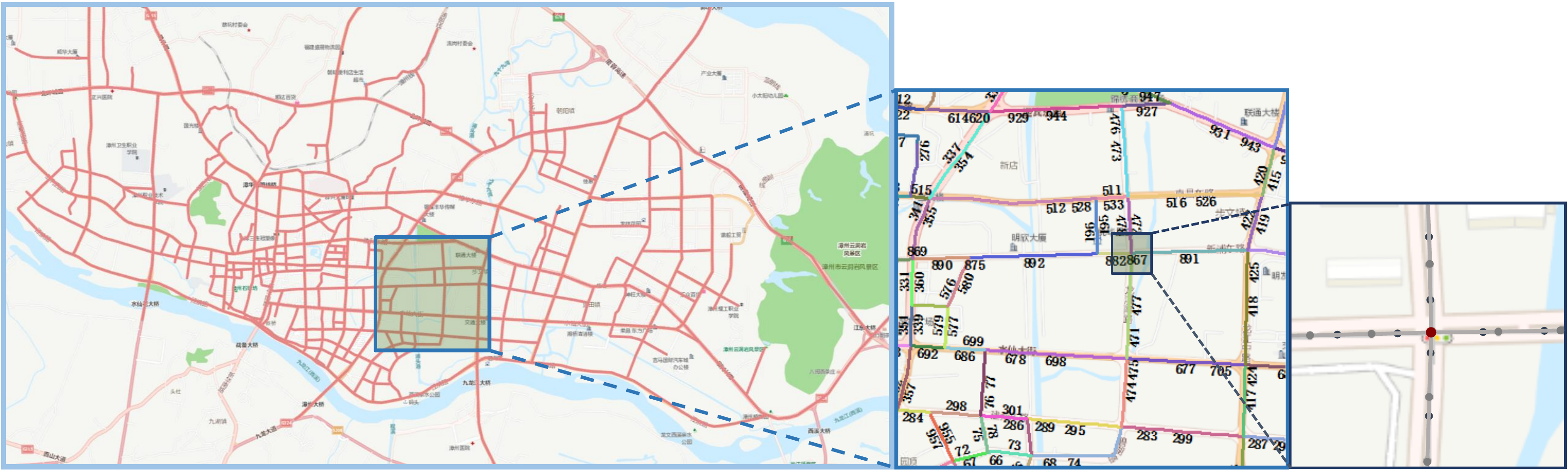}
	\caption{Road network in the urban area of Zhangzhou, China.}
	\label{F:graph}
\end{figure*}

The test GNSS data used in our experiment were collected by 5826 taxis in Zhangzhou, China from April 23rd to May 20th, 2017. The structure of road network is shown in Fig. \ref{F:graph}, in which there are 309 intersection nodes and 1091 links (split into 9886 edges with split length $\delta = 50$ m). 

\begin{figure}[h!]
	\centering
	\includegraphics[width=0.7\linewidth]{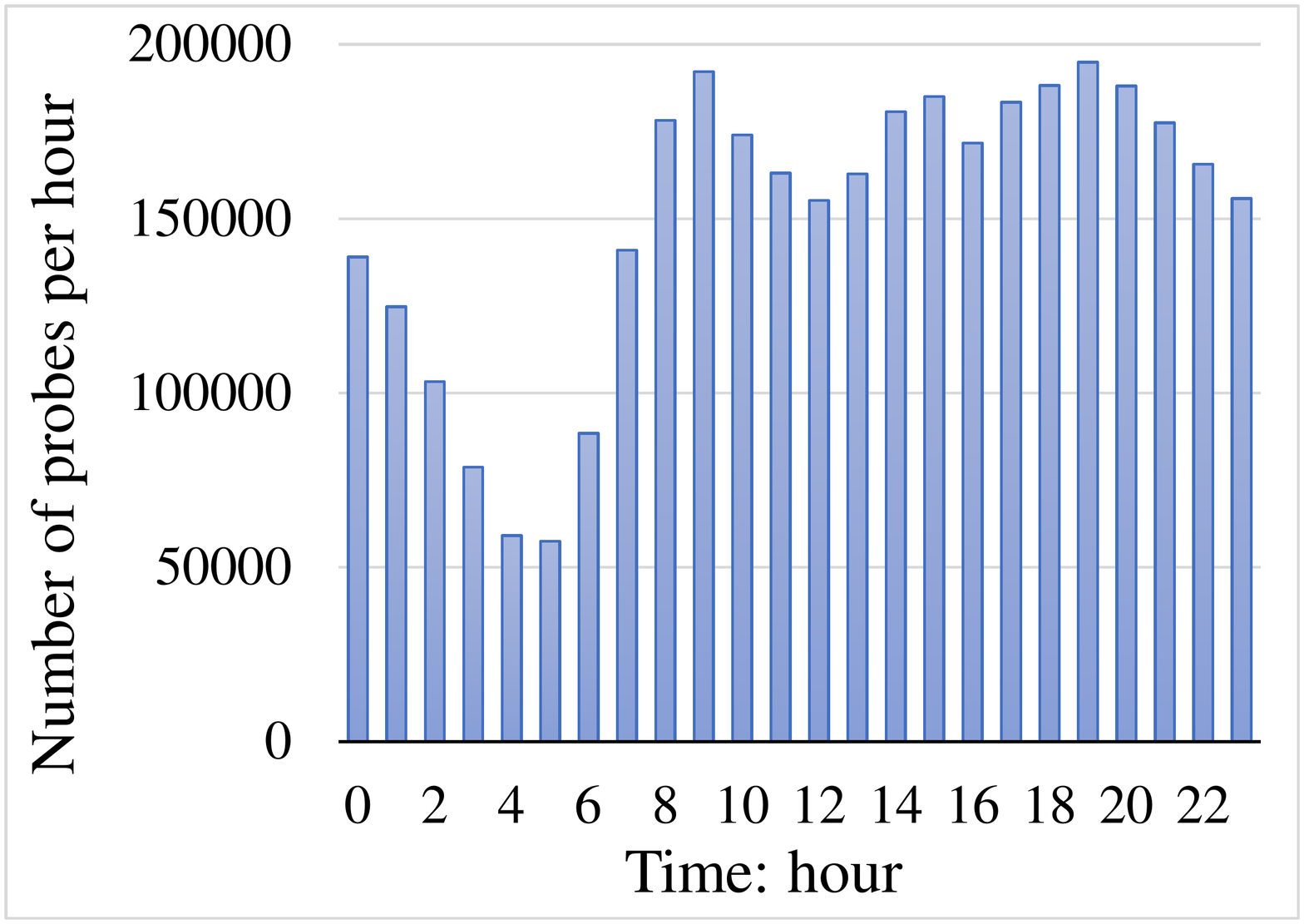}
	\caption{Time distribution of original data.}
	\label{F:distribution}
\end{figure}

We use about 70,000,000 raw GNSS probes with PI $\Delta t = 15$ s in the experiment. Its time distribution is shown in Fig. \ref{F:distribution}. From 0:00 to 7:59, the demand is low, hence many taxis stop and shut down. From 8:00 to 23:59, the demand is high, and the main probe data is collected during this period.

\begin{figure}[ht!] 
	\centering
	\subfloat[][$\Delta t = 15$ s]{\resizebox{0.24\textwidth}{!}{
			\includegraphics[width=\textwidth]{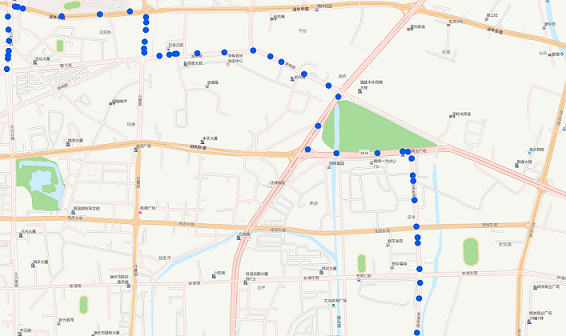}}
		\label{F:I}} 
	\subfloat[][$\Delta t = 60$ s]{\resizebox{0.24\textwidth}{!}{
			\includegraphics[width=\textwidth]{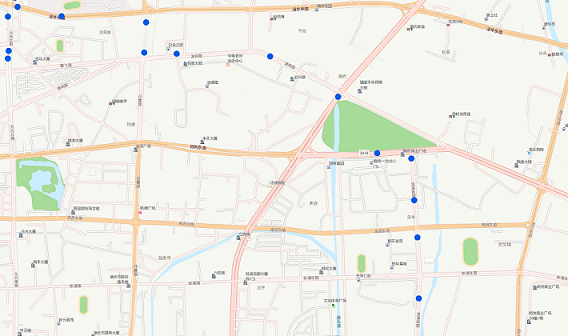}}
		\label{F:II}}
	
	\subfloat[][$\Delta t = 120$ s]{\resizebox{0.24\textwidth}{!}{
			\includegraphics[width=\textwidth]{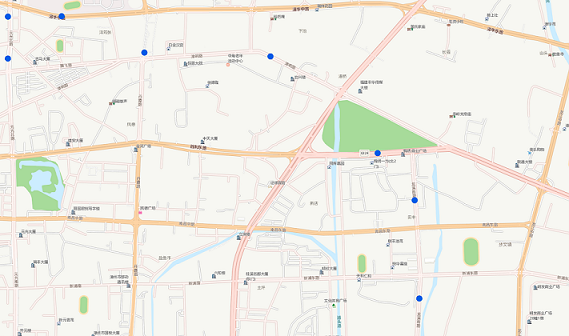}}
		\label{F:III}} 	
	\subfloat[][$\Delta t = 240$ s]{\resizebox{0.24\textwidth}{!}{
			\includegraphics[width=\textwidth]{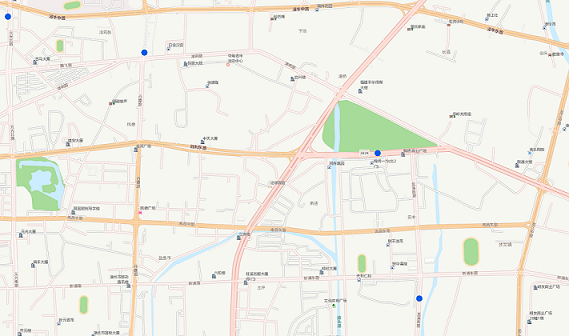}} 		
		\label{F:IV}}
	\caption{Example of a trajectory raw data with different PI ($\Delta t$).}
	\label{F:PI}
\end{figure}

Based on the original data set with 15-s PI, several data sets with sample intervals ranging
from 30 to 300 s were created by removing certain GNSS points from the original data set (integral multiples of 15 s), in order to comparatively test the performance of our algorithm under different data quality.
%%Based on the 15 s raw dataset, we can artificially generate new datasets with various intervals (integral multiples). For example, if the raw data is at: 4s, 19s, 34s, 49s, $\cdots$, we can keep data every 30s interval and obtain: $\Delta t = 30, 60, 120, 180, 240, 300$ s. 
Fig. \ref{F:PI} shows a trajectory sample with various PIs. The shortest-path results with $\Delta t = 15$ s are regarded as the real paths, and we validate and comparatively analyze our algorithm using the lower frequency datasets. It is worth mentioning that there are usually three methods to collect research data for MM. The first method is installing both high-frequency and low-frequency GNSS devices on the same test vehicle and then collecting both types of data from the same trajectories (e.g., \cite{quddus2015shortest}). The second method is directly using high-frequency data to generate low-frequency data by removing certain GNSS points (e.g., \cite{wu2020map,song2018hidden}). The third method is also based on high-frequency data and randomly creates low-frequency data following some distributions (e.g., Normal distribution \cite{rahmani2013path}). We choose the second method because the first method cannot generate lots of data for our study, and the data in the third method may differ greatly from the real data.

Table \ref{T:coef} lists the values of other coefficients we used in the experiment. In addition, our device uses a computer with i7-6700 CPU, 32GB RAM, and 3070ti GPU. 
\begin{table}[h!]
	\caption{Values of coefficients}
	\setlength{\extrarowheight}{2pt}
	\setlength\tabcolsep{8pt}
		\label{T:coef}
	\begin{tabular}{c|c}
		\hline
		Coefficients                                     & Values in experiment  \\ \hline
		max number of candidate paths, $K$               & $\max\big\{0.3\Delta t-18,6\big\}$ \\
		max time steps in SGMN, $k_{\mathrm{\max}}$      & from 1 to 12  \\
		prediction update interval, $\Delta \tau$                     & 300 s    \\
		probe's vicinity radius, $R$                     & 170 m                              \\
		spatial radius in collaboration, $r_\mathrm{s}$  & 300 m (average neighbor is 1)                         \\
		speed coefficient, $\lambda$                     & 0.1                                \\
		temporal radius in collaboration, $r_\mathrm{t}$ & 5 s                                \\                   
		weight in collaboration, $w_\mathrm{c}$          & 0 or 1                                  \\ \hline
	\end{tabular}
\end{table}

\subsection{Comparative analysis of different NN models}

This paper takes SGMN to calculate A-score. However, there are also other NN methods. In this part, we investigated three NN models and calculate their final mean square error $\mathcal{E}$ in Eq. \eqref{E:Error}. Except for SGMN, we pick the Long Short-term Memory model (LSTM) and Graph Markov Neutral Network (GMN) for comparison.
LSTM is a RNN model that could well capture the temporal relationship but without consideration of the spatial neighbors \cite{ma2015long}. GMN is a spatial method in graph convolutional neural network model considering both higher-order spatial neighbors and higher-order temporal neighbors \cite{cui2020graph}. 

The batch size of the training samples is 64. We test NN models with different max time steps ($k_\mathrm{max} = 1,2,\cdots, 12$). The Adam optimization method is adopted to update parameters. We use an early stopping mechanism to avoid over-fitting. The initial learning rate of all models is 0.001. The learning rate is reduced in the order of magnitude until it reaches 0.00001 if there is no improvement in 4 consecutive epochs. The dataset are divided into training, validation, and testing parts according to a 6:2:2 ratio.

\begin{table}[h!]
	\caption{Mean absolute percentage error of the link selection probability for different NN models}
	\setlength{\extrarowheight}{2.5pt}
	\setlength\tabcolsep{16pt}
	\label{T:error}
	\begin{tabular}{c|ccc}
		\hline
		\multicolumn{1}{l|}{Max steps, $k_{\mathrm{\max}}$} & SGMN      & GMN         & LSTM             \\ \hline
1                                 & 4.35\%          & 5.02\%                  & -					     \\
2                                 & 3.90\%          & 4.92\%                  & \textbf{3.14\%}          \\
3                                 & 3.70\%          & 4.94\%                  & 3.25\%                   \\
4                                 & 3.55\%          & 4.89\%                  & 3.30\%                   \\
5                                 & 3.41\%          & 4.97\%                  & 3.21\%                   \\
6                                 & 3.24\%          & 4.70\%                  & 3.37\%                   \\
7                                 & 3.20\%          & 4.83\%                  & 3.20\%                   \\
8                                 & 3.19\%          & 4.88\%                  & 3.21\%                   \\
9                                 & 3.18\%          & 4.92\%                  & 3.18\%                   \\
10                                & 3.16\%          & 4.97\%                  & 3.20\%                   \\
11                                & 3.13\%          & 4.84\%                  & 3.17\%                   \\
12                                & \textbf{3.08\%} & \textbf{4.86\%}         & 3.21\%                   \\
		\hline
	\end{tabular}
\end{table}

We show the results in Table \ref{T:error}. In general, the SGMN can reach the minimum error when $k_\mathrm{max} = 12$, and SGMN is better than LSTM and GMN in respectively best cases. Hence, we set max time steps to be 12 in this paper.

\subsection{Calibration of three weights in final-score}

When calibrating the three weights for different scores, we use the GNSS probes with $\Delta t = 120$ s as the training dataset. After each epoch training, we soft-maximize the $W$. Hence all elements of $W$ are non-negative value and are sumed to one.

When the early stop mechanism was triggered, the number of epochs is 450,000, $\mathcal{E}' = 0.186$, and $W = [0.191,0.521,0.297]$. Hence, we set $[W_\mathrm{P},W_\mathrm{C},W_\mathrm{A}]=[0.2,0.5,0.3]$ in this experiment.

\subsection{Ablation study and comparative analysis}%Comparative analysis of proposed algorithm and other MM algorithms

Because the data before 8:00 a.m., May 20th, 2017 has been used for calibration and validation before, we selected new data from 8:00 a.m. to 8:10 a.m., May 20th, 2017 for analysis in this part, which includes 100 taxis and 4,295 GNSS points. Note that this data is collected from a new time period, hence it is different from the datasets for training, validation, and testing and does not need the cross validation.

There are two performance indices chosen in the analysis. The first index is ``accuracy'', which indicates the percentage of correct probe matching results. The second index is ``recall'', which indicates an average overlap rate in edges between all inferred paths and real paths (e.g., an inferred path with only 50\% correct edges has an overlap rate 50\%). ``accuracy'' and 
``recall'' are complementary. A false path inference result may also lead to a correct point matching; a false point matching result may correctly infer most edges except for the last few edges. In addition, we also attach the ``cost'' index for each experiment, which is calculated by total time cost over the number of matching trajectories.

\begin{table}[h!]
\caption{Performance comparison of different map-matching algorithms}
\centering
\setlength\tabcolsep{1.6pt}
\setlength{\extrarowheight}{3pt}
\label{T:result}	
	\begin{tabular}{cc|ccccccc}
		\hline
		\multicolumn{2}{c|}{\multirow{3}{*}{\begin{tabular}[c]{@{}c@{}}Performance\\ indices\end{tabular}}} & \multicolumn{5}{c}{Proposed method}                                                              & \multirow{3}{*}{\begin{tabular}[c]{@{}c@{}}Quddus'\\ method\end{tabular}} & \multirow{3}{*}{\begin{tabular}[c]{@{}c@{}}Song's\\ method\end{tabular}} \\ \cline{3-7}
		\multicolumn{2}{c|}{}                                                                               & P                    & P+C,             & C+A,             & P+C+A,           & P+C+A,           &                                                                           &                                                                          \\
		\multicolumn{2}{c|}{}                                                                               & \multicolumn{1}{l}{} & $w_\mathrm{c}=1$ & $w_\mathrm{c}=1$ & $w_\mathrm{c}=0$ & $w_\mathrm{c}=1$ &                                                                           &                                                                          \\ \hline
		\multicolumn{1}{c|}{\multirow{3}{*}{30s}}                         & accuracy                        & $94.9\%$       & 96.8\%           & $97.1\%$   & 97.1\%           & \textbf{97.7\%}  & 96.3\%                                                                    & 85.4\%                                                                   \\
		\multicolumn{1}{c|}{}                                             & recall                          & $94.2\%$       & 96.7\%           & $94.8\%$   & 96.8\%           & \textbf{97.4\%}  & 95.9\%                                                                    & 85.2\%                                                                   \\
		\multicolumn{1}{c|}{}                                             & cost, s                         & $0.026$        & 0.059            & $0.051$    & 0.045            & 0.063            & 0.022                                                                     & 0.105                                                                    \\ \hline
		\multicolumn{1}{c|}{\multirow{3}{*}{60s}}                         & accuracy                        & $81.9\%$       & 90.1\%           & $95.2\%$   & 95.8\%           & 95.9\%           & \textbf{96.3\%}                                                           & 71.7\%                                                                   \\
		\multicolumn{1}{c|}{}                                             & recall                          & $89.5\%$       & 92.0\%           & $93.3\%$   & 95.3\%           & \textbf{95.6\%}  & 95.5\%                                                                    & 71.8\%                                                                   \\
		\multicolumn{1}{c|}{}                                             & cost, s                         & $0.062$        & 0.134            & $0.125$    & 0.108            & 0.138            & 0.065                                                                     & 0.190                                                                    \\ \hline
		\multicolumn{1}{c|}{\multirow{3}{*}{120s}}                        & accuracy                        & $64.1\%$       & 80.6\%           & $81.7\%$   & \textbf{91.1\%}  & 90.9\%           & 84.9\%                                                                    & 65.1\%                                                                   \\
		\multicolumn{1}{c|}{}                                             & recall                          & $81.4\%$       & 88.2\%           & $90.5\%$   & \textbf{93.8\%}  & 92.5\%           & 86.1\%                                                                    & 68.5\%                                                                   \\
		\multicolumn{1}{c|}{}                                             & cost, s                         & $0.145$        & 0.413            & $0.327$    & 0.345            & 0.611            & 2.1223                                                                    & 3.8405                                                                   \\ \hline
		\multicolumn{1}{c|}{\multirow{3}{*}{180s}}                        & accuracy                        & $47.1\%$       & 61.3\%           & $65.8\%$   & 76.9\%           & \textbf{79.1\%}  & 73.6\%                                                                    & 57.3\%                                                                   \\
		\multicolumn{1}{c|}{}                                             & recall                          & $72.2\%$       & 76.3\%           & $84.7\%$   & 85.0\%           & \textbf{86.2\%}  & 78.7\%                                                                    & 64.3\%                                                                   \\
		\multicolumn{1}{c|}{}                                             & cost, s                         & $0.276$        & 1.010            & $0.790$    & 0.991            & 1.097            & 3.837                                                                     & 4.653                                                                    \\ \hline
		\multicolumn{1}{c|}{\multirow{3}{*}{240s}}                        & accuracy                        & $43.7\%$       & 51.1\%           & $58.4\%$   & 69.3\%           & \textbf{71.5\%}  & 70.2\%                                                                    & 62.5\%                                                                   \\
		\multicolumn{1}{c|}{}                                             & recall                          & $64.6\%$       & 66.6\%           & $76.5\%$   & 78.3\%           & \textbf{78.6\%}  & 75.8\%                                                                    & 68.4\%                                                                   \\
		\multicolumn{1}{c|}{}                                             & cost, s                         & $0.625$        & 3.341            & $2.836$    & 2.818            & 3.599            & 9.326                                                                     & 23.287                                                                   \\ \hline
		\multicolumn{1}{c|}{\multirow{3}{*}{300s}}                        & accuracy                        & $44.6\%$       & 48.3\%           & $45.5\%$   & 56.2\%           & \textbf{62.6\%}  & 50.6\%                                                                    & 44.2\%                                                                   \\
		\multicolumn{1}{c|}{}                                             & recall                          & $66.7\%$       & 69.4\%           & $75.7\%$   & 72.6\%           & \textbf{76.6\%}  & 60.2\%                                                                    & 52.1\%                                                                   \\
		\multicolumn{1}{c|}{}                                             & cost, s                         & $1.331$        & 7.422            & $5.138$    & 5.934            & 7.584            & 16.136                                                                    & 40.147                                                                   \\ \hline
	\end{tabular}
\end{table}

We test our algorithm with an ablation study, and the results are shown in Table \ref{T:result} (Proposed method). There are six different PIs ($\Delta t = 30, 60, 120, 180, 240, 300$ s) and five types of combinations in which P, C, and A mean the P-score, C-score, and A-score, respectively. Comparing the C+A and P+C+A ($w_c = 1$), we can find that P is effective to increase both the accuracy and recall. A comparison of P and P+C shows that C is effective to increase both the accuracy and recall. A comparison of P+C and P+C+A ($w_c=1$) illustrates that A is effective to increase both the accuracy and recall. Finally, Comparing the $w_c=0$ and $w_c=1$ for P+C+A, we can find that the coefficient $w_c=1$ performs better in most PIs except for $\Delta t = 120$ s, hence we should let $w_c =1$ in the algorithm. In general, as traditional methods mainly use the present data, the collaborative data (including historical data) can further improve MM's precision especially when PI is from 60 to 180 s. Real-time traffic state is also useful for enhancing MM's results, especially when $\Delta t >30$ s.

Table \ref{T:result} also attaches two methods (Quddus' and Song's) for comparison. Quddus' method mainly considered the classical shortest-path algorithm within a circle region\cite{quddus2015shortest}. Song's method further considered historical data using hidden Markov model \cite{song2018hidden} within a circle region, and it ignores the bearing and direction information. Their methods are classical in their corresponding areas (present data and present+historical data), and our algorithm is also inspired by both the methods. Hence, we compare them with our algorithm in this experiment. 
In general, the proposed method (P+C+A with $w_c=1$) outperforms Quddus' and Song's methods, especially when $\Delta t >60$ s. That means our method %It means that our method 
performs better if probing frequency is lower. % compared with other MM algorithms. 
Only when $\Delta t = 60$ is Quddus' accuracy higher than our method. This may come from samples' random error because Quddus' accuracy from $\Delta t = 30$ s to $\Delta t = 60$ s keeps the same (96.3\%). Furthermore, both Quddus' method and our method are much better than Song's method. Song et al. used 5 months historical data to achieve the best accuracy \cite{song2018hidden}, but we only have 4-week data. That may be the main reason why it performs unsatisfactorily. 

Although our algorithm uses various kinds of data, much of the work is one-off, and the processed historical results can be stored and reused in the future. Hence it is not time costly even when $\Delta t>60$ s. Actually, it is the candidate paths searching work that occupies the main calculation time, especially when $\Delta t >30$ s. In general, %Quddus'
traditional MM methods with the simplest rule cost the least time when $\Delta t = 30$ or $60$ s, but their ``cost'' increases dramatically if the probing frequency is lower. When $\Delta t>60$ s, our ``cost'' is smaller. %method performs the best in  ``cost''. 
That is because our modified top-K shortest-path method excludes unreasonable trajectories using a smaller graph (within an ellipse region). Instead, Quddus' and Song's methods search the path in a larger graph (within a circle region) with more choices. This effect is more significant with longer distance between two probes.

\begin{figure}[h!]
	\centering
	\includegraphics[width=0.9\linewidth]{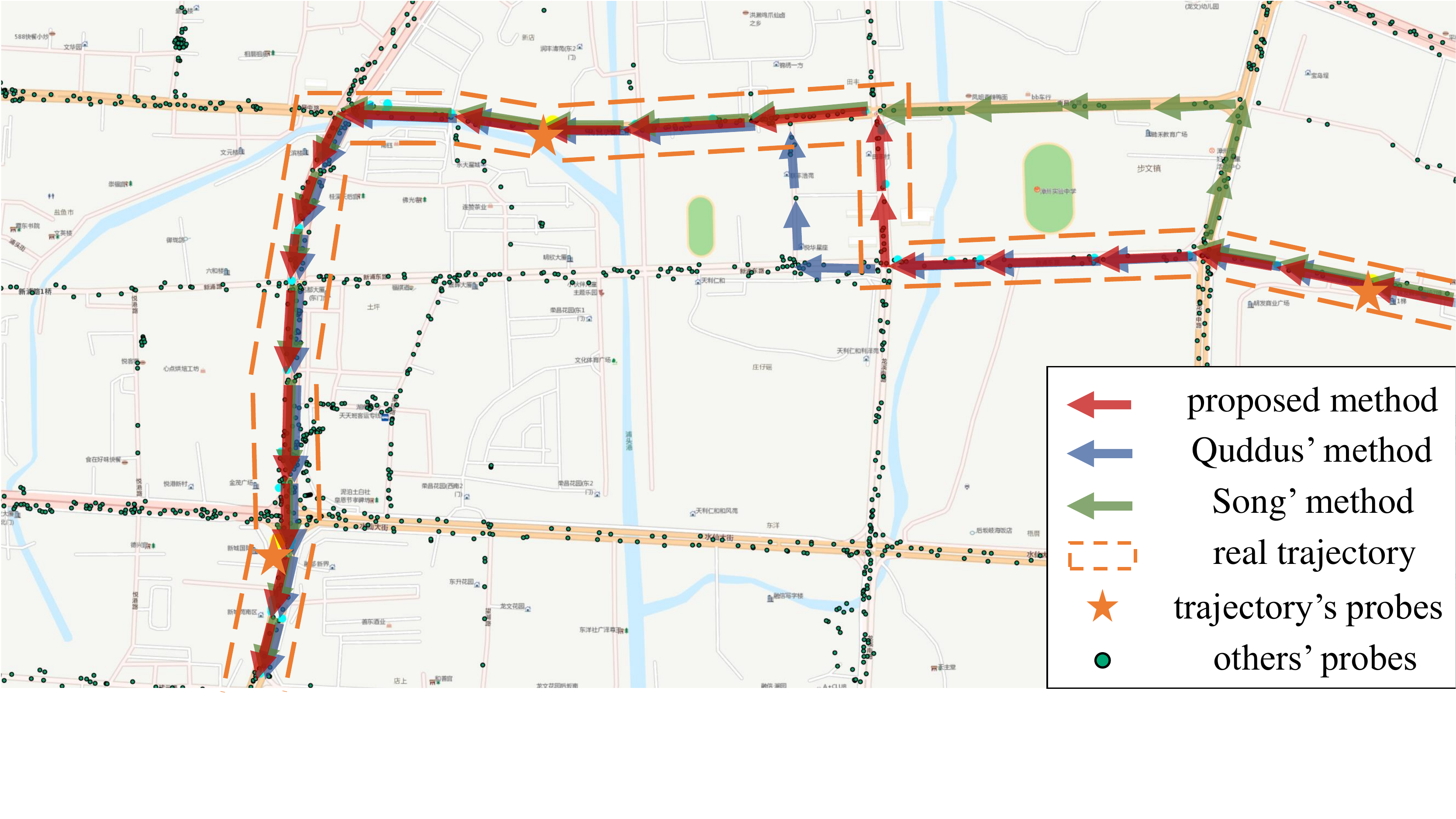}
	\caption{A sample of trajectory for three MM algorithms ($\Delta t= 240$ s).}
	\label{F:sample_MM}
\end{figure}

Finally,  Fig. \ref{F:sample_MM} shows an example of three MM results, with which we can clearly find their differences in prediction. When moving north, Quddus' method chooses the left path, our method chooses the middle path (real path), and Song's method choose the right path. Compared with the correct path, Quddus' method chooses a path with small volume because it does not consider the real-time traffic state. Besides, Song's method ignores the vehicle's heading information and chooses the right path with biased direction.

\section{Conclusion}
%三层 
Limited by the current technologies, further improving the map-matching (MM) accuracy is challenging work, especially for the low-frequency probing data. With the popularity of GNSS devices, the operators or government can obtain ``Big'' data in a long-term operation, which is useful for traffic research, including the MM. Most of the existing MM studies focus on vehicles' own present data (P-data), and a few of MM studies took the historical data into consideration. To our best knowledge, there is no study combining MM with the real-time traffic state estimation. 

Different data may all contribute to the accuracy of present trip's MM to some degree, but %their concentrations of value can be quite different
it is hard to extract valuable information (to present trip) from other trips. To make full use of ``Big'' data, we split all data (and historical MM results) into 4 groups, and design three scoring methods to ``extract'' helpful information. 
\underline{First}, all data are split into P-data: present data, C-data: collaborative historical data, A-data: all data in network in recent period, and useless data (Section \ref{S:Intro}: Introduction). \underline{Second}, we design three ``judges'' including P-score, C-score and A-score to use the three groups of useful data, respectively (Section \ref{S:SM}: Scoring Methods). For any candidate path, P-score evaluates it from aspects of speed and bearing, C-score evaluates it considering its historical usage rate in collaborative group, and A-score evaluates it based on the prediction of current traffic distribution (obtained through spectral graph Markov neutral network). Then, the algorithm calculates the weighted mean score of the three scores (whose weights could be calibrated using a simple neutral network structure). \underline{Third}, we design a modified top-K loopless shortest-path method to search the candidate paths (from P-data) within an ellipse sub-graph and then infer the path (projected location) based on the given scoring methods (Section \ref{S:PI}: Path Inference). 

We use 70 million taxi GNSS data (4 weeks) from a city of China to calibrate and test our method in experiment. The results show that, every ``judge'' helps to enhance MM accuracy, and our method performs better than the others (\cite{quddus2015shortest,song2018hidden}) in the indices of ``accuracy'' (percentage of correct probe-matchings), ``recall'' (mean overlap rate in path inference) and average ``cost'' (when probing frequency is smaller than 1/60 HZ). Furthermore, The advantage of our method becomes larger if the probing frequency is lower. 

However, limited by the datasets, our ``real'' trajectories are based on 15-s datasets, whose frequency is still too small and may lead to some errors in getting the ``real'' ones (but this should not influence the overall trend shown in Table \ref{T:result} because the predicted trajectory from 15-s data serves as a baseline with random errors for all experiments.). Besides, our method used real-time information from traffic state estimation, which makes our method hard to be used online or in real time (because the calculation and data transmission may delay the process). One way to deal with it is using previous traffic state information, hence procession of the historical results can be completed in advance. But this method will slightly reduce the prediction accuracy.  
In addition, if the data's amount is not enough, the C-score and A-score may work poorly, and our algorithm will degenerate into a method with only P-score (similar to traditional MM). To implement our work in the real world, we need to not only guarantee enough data amount (at least one-month network data as in this paper) but also improve the algorithm's searching efficiency.  
	 In the future, we try to change the design of different ``judges'' to further improve the MM's efficiency and precision, and the proposed method should be tested in an original datasets with higher probing frequency. It would also be meaningful to test our algorithm with non-artificial data in the future if available.

\ifCLASSOPTIONcaptionsoff
  \newpage
\fi

\bibliographystyle{IEEEtran}

\bibliography{literature}

\vspace*{-0.6cm}
\begin{IEEEbiography}[{\includegraphics[width=1in,height=1.25in,clip,keepaspectratio]{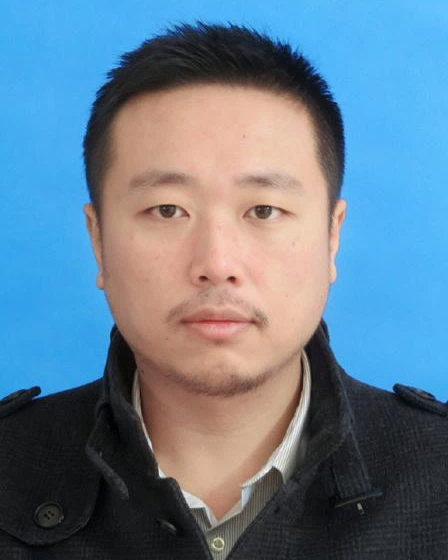}}]{Jie Fang} 
	received his M.S. and Ph.D. degrees in the Civil and Environmental Engineering Department from the University of Wisconsin-Madison, U.S... He is currently a Professor in the Department of Civil Engineering, Fuzhou University, and the director of their transportation program. His research interests include intelligent transportation systems, proactive traffic control, and the utilizing of artificial intelligence algorithms in the transportation area.
\end{IEEEbiography}

\vspace*{-0.6cm}
\begin{IEEEbiography}[{\includegraphics[width=1in,height=1.25in,clip,keepaspectratio]{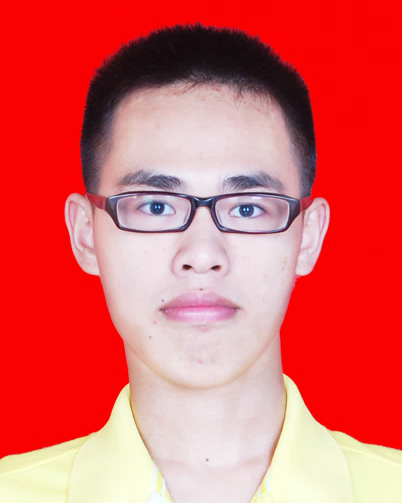}}]{Xiongwei Wu} 
	received a bachelor’s degree in Transportation Engineering from the College of Civil Engineering, Fuzhou University, China, in 2020, where he is currently pursuing a master’s degree in the College of Civil Engineering. His research interests include map matching, spatial-temporal data mining, and their application to ITS.
\end{IEEEbiography}

\vspace*{-0.6cm}
\begin{IEEEbiography}[{\includegraphics[width=1in,height=1.25in,clip,keepaspectratio]{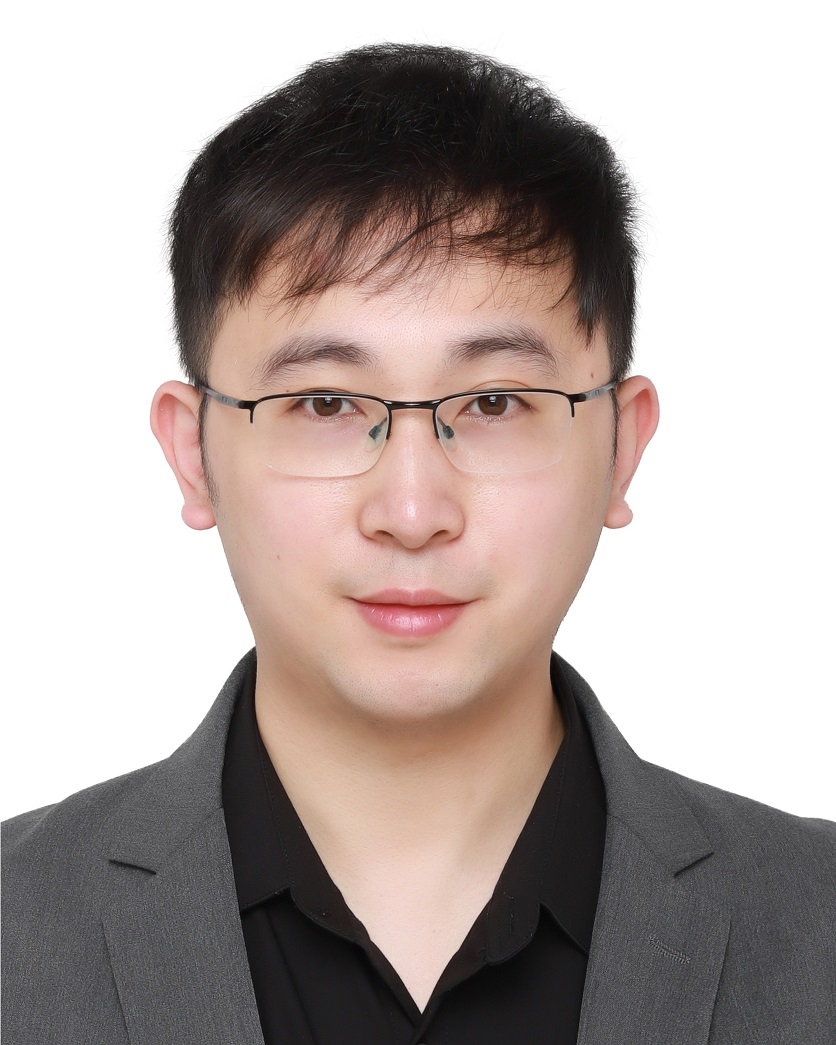}}]{DianChao Lin}
	was born in Fujian, China, in 1990. He received his B.Sc. and M.Sc. degrees in Traffic Engineering and Traffic Information Engineering \& Control from Tongji University, Shanghai, China, in 2013 and 2016, respectively, and the Ph.D. degree in Transportation Planning \& Engineering from New York University, NY, U.S.A., in 2021. He is currently an assistant professor with the School of Economics and Management, Fuzhou University, Fuzhou, China. His research interests include traffic management with economic schemes, connected vehicles, applications of game theory to automated vehicles, non-motorized traffic behavior and multi-modal traffic flow.
\end{IEEEbiography}

\vspace*{-0.6cm}
\begin{IEEEbiography}[{\includegraphics[width=1in,height=1.25in,clip,keepaspectratio]{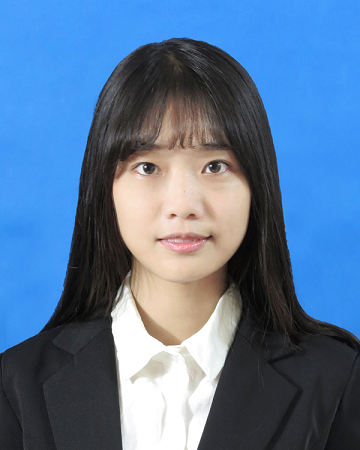}}]{Mengyun Xu}
	received the B.S. degree and the M.S. degree in Transportation Engineering from the department of civil Engineering at Fuzhou University, Fuzhou, China, in 2017 and 2020. She is currently pursuing a Ph.D. degree in the Intelligent Transport System Research Center, Wuhan University of Technology. Her research interests include Intelligent Transportation Systems, traffic data mining, and Artificial Intelligence.
\end{IEEEbiography}

\vspace*{-0.6cm}
\begin{IEEEbiography}[{\includegraphics[width=1in,height=1.25in,clip,keepaspectratio]{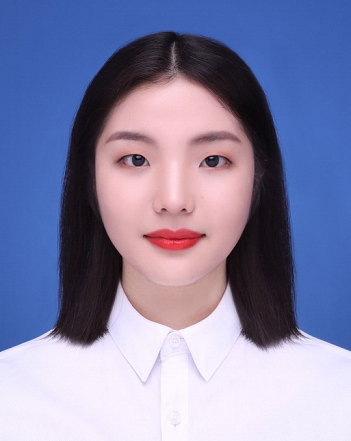}}]{Huahua Wu}
	received a bachelor’s degree from the School of Transportation, Fujian University of Technology, China, in 2020. She is pursuing a master’s degree from the College of Civil Engineering, Fuzhou University. Her research interests include neural networks and the application of big data in transportation systems. 
\end{IEEEbiography}

\vspace*{-0.6cm}
\begin{IEEEbiography}[{\includegraphics[width=1in,height=1.25in,clip,keepaspectratio]{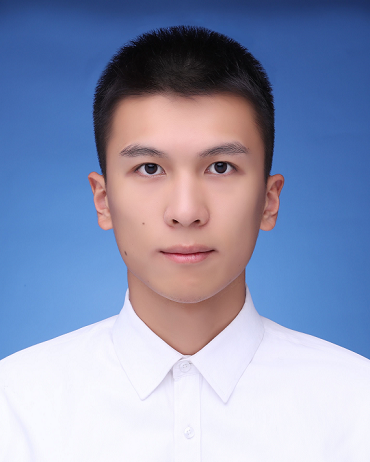}}]{Xuesong Wu}
	received a bachelor’s degree from the College of Civil Engineering, Fuzhou Agriculture and Forestry University, China, in 2020. He is pursuing a master’s degree from the College of Civil Engineering, Fuzhou University. His research interests include machine learning and intelligent transportation systems.
\end{IEEEbiography}

\vspace*{-0.6cm}
\begin{IEEEbiography}[{\includegraphics[width=1in,height=1.25in,clip,keepaspectratio]{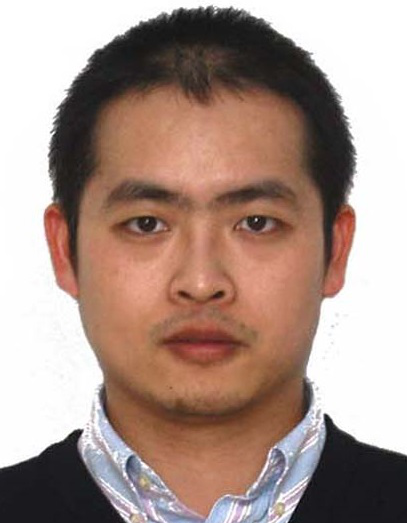}}]{Ting Bi}
	received the B.Eng. degree in Software Engineering from Wuhan University, China in 2010, and received the M.Eng. and PhD degrees in Telecommunications from Dublin City University, Ireland in 2011 and 2017, respectively. He is currently a Assistant Professor in the Department of Computer Science, Maynooth University. His research interests include mobile and wireless communications, multimedia and multi-sensory media streaming over wireless access networks, user quality of experience, and energy saving for mobile devices. He is a member of IEEE, ACM, RDA. 
\end{IEEEbiography}

\end{document}